\documentclass[%
floatfix, 
reprint,
]{revtex4-2}
\usepackage{amsmath, amssymb, amsthm,amsfonts, bm} 
\usepackage[english]{babel}
\usepackage{bbm}
\usepackage{graphicx}
\usepackage{url}
\usepackage{epstopdf}
\usepackage[a4paper,bindingoffset=0.5cm,left=2cm,right=2cm,top=2.5cm,bottom=2cm,footskip=.8cm]{geometry}
\usepackage[pdfencoding=auto]{hyperref}
\usepackage{kbordermatrix}
\usepackage{tikz}
\usetikzlibrary{arrows, automata,positioning,calc,shapes,decorations.pathreplacing,decorations.markings,shapes.misc,petri,topaths}
  \usepackage{pgfplots}
  \pgfplotsset{compat=newest}
  \usetikzlibrary{plotmarks}
  \usepackage{grffile}
\newlength\figureheight
  \newlength\figurewidth
\pgfplotsset{%
    tick label style={font=\scriptsize},
    label style={font=\footnotesize},
    legend style={font=\footnotesize},
         every axis plot/.append style={very thick}
}
 
\usepackage{rotating}
\usepackage{amsbsy,enumerate}
\usepackage{graphicx}
\usepackage{ccaption}
\usepackage{comment}
\usepackage{mathrsfs} 
\usepackage{mathtools}
\usepackage{xcolor}
\usepackage{algorithm}
\usepackage{algpseudocode}
\usepackage{subfig}

\usepackage[normalem]{ulem}

\makeatletter
\newcommand{\specialcell}[1]{\ifmeasuring@#1\else\omit$\displaystyle#1$\ignorespaces\fi}
\makeatother

\newcommand{\vb}{\vspace{3.2mm}}

\allowdisplaybreaks

\newcommand{\belief}[1]{\bm{x}(#1)}

\theoremstyle{plain}
\newtheorem{theorem}{Theorem}

\newtheorem{lemma}{Lemma}

\newtheorem{assumption}{Assumption}
\theoremstyle{definition}

\newtheorem{conjecture}{Conjecture}
\newtheorem{remark}{Remark}

\begin{document}

\title[Intra-population trust]{Interpersonal trust: Asymptotic analysis of a stochastic coordination game with multi-agent learning}

\author{Benedikt V. Meylahn, Arnoud V. den Boer, and Michel Mandjes}

\begin{abstract}
\noindent{\sc Abstract} We study the interpersonal trust of a population of agents, asking whether chance may decide if a population ends up with high trust or low trust. We model this by a discrete time, stochastic coordination game with pairwise interactions occurring at random in a finite population. Agents learn about the behavior of the population using a weighted average of what they have observed in past interactions. This learning rule, called `exponential moving average', has one parameter which determines the weight of the most recent observation and may thus be interpreted as the agent's memory. We prove analytically that in the long run the whole population always either trusts or doubts with probability one. This remains true when the the expectation of the dynamics would indicate otherwise. By simulation we study the impact of the distribution of the payoff matrix and of the memory of the agents. We find, that as the agent memory increases (i.e., the most recent observation weighs less), the actual dynamics increasingly resemble the expectation of the process. We conclude that it is possible that a population may converge upon high or low trust between its citizens simply by chance, though the game parameters (context of the society) may be quite telling.
\vb

\noindent
{\sc AMS Subject Classification (MSC2020).} Primary: 91-10 (Mathematical modeling or simulation for problems pertaining to game theory, economics, and finance), 91A22 (Evolutionary games); Secondary: 91D15 (Social learning)
\vb

\noindent
{\sc PACS number(s).} 02.50.Le (Decision theory and game theory), 87.23.Ge (Dynamics of social system), 89.75.Fb (Structures and organization in complex systems)
\vb

\noindent
{\sc Affiliations.} BV Meylahn, M Mandjes and AV den Boer are all affiliated with: Korteweg-de Vries Institute for Mathematics, University of Amsterdam; Science Park 904, 1098 XH Amsterdam; The Netherlands ({\it contact}: {\tt\scriptsize  b.v.meylahn[at]uva.nl}).
M Mandjes is also affiliated with: Mathematical Institute, Leiden University, Niels Bohrweg 1, 2333 CA Leiden; The Netherlands.
\vb

\noindent Version: \today. 
\end{abstract}

\maketitle
\textbf{We are interested in a finite population playing a stochastic coordination game in random pairwise interactions which we use to model an interaction requiring trust. Consider an agent's point of view: `If others are abusing trust I prefer not to act trustingly while if others are honorable, then I would prefer to trust them.' Agents ideally coordinate on either trusting/acting honorably or on distrusting/abusing trust. The agents in our model are endowed with an adaptive learning rule. Models of learning in evolutionary game theory often assume that the agents' opponent(s) are playing a fixed strategy and aim to learn this from experience. This makes little sense when all the agents are employing that learning rule (and thus updating their strategy). Our learning rule instead is an exponentially weighted moving average of the agents observations (also called simple exponential smoothing) which does not assume that the opponents' strategy is fixed. Our model of a stochastic coordination game does away with fixed payoffs with noise, and instead allows for a very broad category of random variables as payoffs with very mild conditions on their distributions. This is a new approach in the context of agents who learn based on experience, though common in the study of the replicator equation. Another distinguishing element is that the stochastic game is dynamic as the payoffs are drawn at random anew at each time step. We prove convergence of belief and behavior in the long run to pure Nash equilibria (always: trust/act honorably, or distrust/abuse trust). We conclude with simulations to explore the relationship between model parameters and relative probability of convergence to the trustful steady state and the rate of convergence. We see the surprising result that a shorter agent memory (i.e., more weight to the most recent observation) dampens the effect of the payoff distributions on the model outcomes.}
\section{Introduction}

Trust is beneficial to a societies' functioning yet there is not a uniform amount of trust in different societies across the world~\cite{owidtrust}. Is there an inherent difference between the societies which exhibit high trust and those that exhibit low trust? In Denmark, it is common practice to leave sleeping babies in their prams outside on the pavement, while its parents are shopping inside~\cite{Thomsen2006}. In contrast, one need only look around and observe widespread electric fences and infrared alarm systems in a typical South African neighborhood to see that its citizens do not believe each other to be trustworthy.

What are the antecedents of low and high trust? Is there some insurmountable difference between the citizens of Denmark and those of South Africa, or is it possible that these differences are matters of chance? Furthermore we ask what effect the structural properties of the interaction have on the probability of low or high trust emerging. 

We apply the tool of evolutionary game theory with agent learning to the question of interpersonal trust. Instead of studying an $N$-person trust game~\cite{Berg1995} as in~\cite{Kumar2020, Sun2022,Feng2023,Liu2023, Wang2024}, we make a simplification. In particular we study a 2-person coordination game played by two random agents drawn repeatedly from a population of $N$ agents. This allows us to study what happens if both agents involved in the interaction learn from the interaction. This approach also illuminates the dynamics that occur for non-coordination games in which agents have positive externalities (utility derived from taking the same action as other players). We consider the problem of placing trust and acting in a trustworthy manner in society as a coordination game. We condense the matter of placing trust and honoring placed trust into the action \textit{trust}. Similarly, not placing trust and abusing others' trust are condensed into the action \textit{doubt}. Agents are happy to act trustingly while others are acting in a trustworthy manner, but prefer to act distrustfully when their neighbors are abusing trust, hence the payoffs follow those of a coordination game. In the model, a population of agents interacts by a random matching of two agents per round. These agents take the action which they believe to maximize their one-round expected utility. This expectation is based on their belief of `the probability that a random other in the population acts trustingly,' and the randomly realized payoff in their stochastic coordination game. In this fictitious play-like~\footnote{Fictitious play describes the agent learning paradigm in which each agent assumes that they are playing against a fixed mixed strategy profile. The agents use historical play to try and learn the profile they believe to be playing against and act myopically optimally against it~\cite{FudenbergLevine1998}. Our agents' actions are akin to fictitious play as they base these myopically on their belief. However, their belief is updated as if their opponents strategy can change.} model of learning, agents update their belief based on the exponential moving average rule.

We prove convergence of behavior and beliefs to the always-trust or always-doubt steady states in the long run. Furthermore, we highlight the impact that the exponential smoothing learning parameter has on the relative probability of convergence to the always-trust versus always-doubt steady state. Surprisingly, a greater learning rate (akin to a shorter memory of the agents) has dampening effect on the impact of the game payoff parameter distributions. Conversely, based on our numerical simulations we conjecture that a learning rate that approaches zero (an infinite memory) may create a phase transition between the always-trust and the always-doubt steady states depending on the distributions of the payoff variables.

Evolutionary game theory is a commonly used tool when it comes to understanding antecedents of social dynamics~\cite{Burger2013,Lim2018,Lim2020,Meylahn2021,Godara2022,Yang2023,Wang2023}. In particular, models with agent learning have been studied extensively~\cite{Fudenberg1993,Hofbauer2002,Kaniovski1995,Tuyls2006,Laird2013,Baudin2022,Godara2023,Meylahn2023A,Lu2023,Zhu2023}. Coordination games with slight payoff perturbations have been studied extensively as a tool to understand norm-formation~\cite{Kaniovski1995,Young1993,Xie2012,Bilancini2020,Sayin2022}. Our model is the first we know of which considers the case with fully stochastic payoffs rather than small random perturbations to a fixed payoff-matrix in the context of multi-agent learning. 

We note the rich history of the fully stochastic payoff-matrix~\cite{Joireman1996, Ert2011,Berg1998, Han2012, Duong2020, Karlin1974,Karlin1975, Zheng2017, Feng2022, Perc2007, Zeng2022, Qian2024}. In particular there is a line of work on the statistics of the replicator (or the replicator-mutator) dynamics under the random draw of the game payoff-matrix~\cite{Han2012, Duong2020}. This is in contrast with~\cite{Karlin1974,Karlin1975, Zheng2017, Feng2022}  which study the replicator dynamics but in a changing environment (i.e., the payoff-matrix is drawn in each generation from some distribution). The replicator equation models a process of selection in which the reproductive fitness of a species (or strategy) is equal to its average payoff in the game. The object of interest in such studies is the number of (mixed) equilibria and their stability. Learning by imitation has also been studied in the context of a stochastic payoff-matrix~\cite{Perc2007, Zeng2022, Qian2024}. In such models agents choose a neighbor and compare their most recent payoffs. If their neighbor has a greater payoff, the agent adopts their neighbors strategy with a given probability. In this branch of the literature one looks for elements in the randomness or the population structure which either promote or inhibit the emergence of trust.

The context of our model is experience based learning, i.e., the agents are learning about the population based on the strategies agents observe their opponents using and use this information to optimize their actions. Examples of this kind of learning with payoff matrices which are either static, or perturbed by small noise can be found in~\cite{Sato2002,Sato2003, Sato2005, Banisch2019}, and~\cite{Young1993,Kaniovski1995} respectively. The fact that agents converge to the always-trust or the always-doubt state complements (is in qualitative agreement with) the existing results of convergence to pure Nash equilibria~\cite{Nash1951} in the context of small perturbations and a decreasing learning rate in~\cite{Fudenberg1993,Kaniovski1995,Benaim1999,Hofbauer2002}. 

Our model illustrates that a \textit{small} population may, by chance, end up in a low or a high trust state. As the size of the population grows however, chance plays less of a role and the outcome is almost entirely decided by the distributions of their payoff parameters.
The distributions of the payoff variables, represent the context that a population is in. In situations where the payoff distributions promote doubting (low trust) it would behoove the individuals to have a shorter memory so that happy mistakes could lead to long-term trust.

Thus because neither South Africa nor Denmark have `small' populations, our model leads us to believe that the low and high trust observed respectively may be largely due to the structural properties of interactions. By this we mean the learning rate or the distributions of the payoffs in the coordination game. Our stylized mathematical model is of a high level of abstraction and aims to capture the essence of the underlying dynamics. Evidently, in the (inherently more complex) real dynamics, there are various other differences between South Africa and Denmark.

In \S\ref{BaseModel} we formulate the model formally, which we follow in \S\ref{Asym_as} with an asymptotic analysis. This culminates in the main theorem regarding convergence of belief and behavior of the agents. In \S\ref{Illustration} we provide an illustration of the model with only two agents which highlights the perhaps surprising nature of our main results. Having identified the asymptotic behavior of the dynamics, we introduce and discuss the results of a simulation study in \S\ref{Simulation}. The simulation serves to elucidate the interdependence of the parameters on the chance of convergence to high or low trust behavior. We conclude in \S\ref{Conclusion} with a discussion of the results and possible future work.

\section{Model}
\label{BaseModel}
We consider a population of $N\in\mathbb{N}$ ($N\geq 2$) agents who engage in a game played repeatedly in discrete rounds indexed by $t\in\mathbb{N}$. $\mathbb{N}$ represents the natural numbers, $1,2,3,\ldots$ in this paper. At the start of each round $t\in \mathbb{N}$, a pair of agents $(I(t),J(t))$ is chosen uniformly at random from the set of tuples $\{(i,j)\in\{1,\ldots,N\}:i\neq j\}$. This may either be interpreted as if the population is fully mixed and there is no population structure, or equivalently as if the structure imposed on the population is the fully connected network. The chosen pair of agents play a $2\times 2$ coordination game. In this game $I(t)$ takes the role of agent $k=1$ (the row agent) and $J(t)$ takes the role of agent $k=-1$ (the column agent). We define $g(k,t)$:
 \begin{equation*}
    g(k,t) := \begin{cases}
        I(t) \quad &\text{if }k=1,\text{ and}\\
        J(t) & \text{if }k=-1,
    \end{cases} \quad \text{for }t\in\mathbb{N}.
\end{equation*}
This allows us to retrieve an agent's index within the population given their role in the game.
The action of agent $k\in\{1,-1\}$ at time $t\in\mathbb{N}$ is denoted $A_k(t)\in\{T,D\}$.
The action $T$ denotes trusting and $D$ denotes the action of doubting. We define the payoff bimatrix of the game as $\Pi(t)$ at time $t\in \mathbb{N}$: 
\begin{equation}\label{Game:Matrix}
\Pi(t)= \kbordermatrix{
~ & T & D \cr
T & (U_1(t),U_{-1}(t)) & (V_1(t),W_{-1}(t)) \cr
D & (W_1(t),V_{-1}(t)) & (Y_1(t),Y_{-1}(t)) \cr}.
\end{equation}
The first value of the matrix entry $\Pi_{l,m}(t)$ is the reward obtained by the row agent when playing the action in row $l$ against a column agent who plays the action in column $m$ during round $t\in\mathbb{N}$. Conversely, the second value of the matrix entry $\Pi_{l,m}(t)$ is the reward obtained by the column agent when playing the action in column $m$ against a row agent who plays the action in row $l$ during round $t\in\mathbb{N}$. We define $\Pi_k(t)$ as the matrix containing only the payoffs to agent $k\in\{-1,1\}$ at time $t\in\mathbb{N}.$

We assume that the payoffs in (\ref{Game:Matrix}) $\{U_k(t): k\in\{1,-1\}, t\in\mathbb{N}\}$ are all independent and identically distributed continuous random variables. The same is true for $\{V_k(t): k\in\{1,-1\}, t\in\mathbb{N}\}$, $\{W_k(t): k\in\{1,-1\}, t\in\mathbb{N}\}$, and $\{Y_k(t): k\in\{1,-1\}, t\in\mathbb{N}\}$. To ensure that the game has the structure of a coordination game we assume that 
\begin{equation}\label{eq:R_restrict}
    U_k(t)>V_k(t) \quad \text{and} \quad U_k(t)>W_k(t) \quad \text{w.p. }1,
\end{equation}
and 
\begin{equation}\label{eq:P_restrict}
    Y_k(t)>V_k(t) \quad \text{and} \quad Y_k(t)>W_k(t) \quad \text{w.p. }1,
\end{equation}
for $k\in \{1,-1\}$ and all $t\in\mathbb{N}$. 
\begin{remark}
The restrictions (\ref{eq:R_restrict}) and (\ref{eq:P_restrict}) are stronger than what we need for our results. Our analysis holds whenever 
\begin{equation*}
Y_k(t)-V_k(t)> 0,
\end{equation*}
\begin{equation*}
     U_k(t)-W_k(t)\geq 0,
\end{equation*}
and 
\begin{equation*}
    U_k(t)-W_k(t)+Y_k(t)-V_k(t)>0,
\end{equation*}
for all $t\in\mathbb{N}, k\in\{1,-1\}$.
We add the assumptions (\ref{eq:R_restrict}) and (\ref{eq:P_restrict}) to show the relevance of the analysis to a coordination game. 
\end{remark}
In our setting, the agents model the population behavior with a belief on the probability that a randomly chosen individual that they play against would trust. Let $x_i(t)$ denote the belief of agent $i\in\{1,\ldots,N\}$ at the beginning of round $t\in\mathbb{N}$ on the likelihood that they will encounter an agent playing trust. The vector holding the beliefs of the agents in the population at time $t$ is: $\belief{t}=(x_1(t),\ldots,x_N(t))^\top\in[0,1]^N$.

At the start of each round each agent $k\in\{1,-1\}$ only observes their own payoffs $U_{k}(t)$, $Y_{k}(t)$, $V_{k}(t)$, $W_{k}(t)$ but not the payoffs of their opponent $U_{-k}(t)$, $Y_{-k}(t)$, $V_{-k}(t)$, $W_{-k}(t)$. Furthermore, the agents are not aware that their opponent's payoffs follow the structure of a coordination game. The agents model their opponent's behavior entirely by their belief on the likelihood that their opponent take the trust action. We define $u^T_{k}(t)$ and $u^D_{k}(t)$ as the expected utility for agent $k\in\{1,-1\}$ playing $T$ and $D$ respectively during round  $t\in\mathbb{N}$ based on agent $k$'s belief:
\begin{align}
    u^T_{k}(t) &:= x_{g(k,t)}(t) U_{k}(t)+ (1-x_{g(k,t)}(t))V_{k}(t),\\
    u^D_{k}(t) &:= x_{g(k,t)}(t) W_{k}(t) + (1-x_{g(k,t)}(t))Y_{k}(t).
\end{align}

Myopic decision making is a common assumption~\cite{BalaGoyal1998,Wang2006,Raghunandan2012,Li2013,Harel2021,Lee2023}. Furthermore, there is experimental work that suggests that humans indeed act at least semi-myopically~\cite{Yu2013,Zhang2013}. As such, we assume the agents take actions myopically:
\begin{assumption}[Myopic rationality]\label{As:Myopic}
    We assume agents to be myopically rational, taking the action which maximizes the 1-round expected utility, with ties favoring trust:
\begin{equation*}
A_{k}(t) := 
    \begin{cases}
        T\quad &\text{if }  u^T_{k}(t)\geq u^D_{k}(t)\\
        D&\text{else,}
    \end{cases}
\end{equation*}
for $t\in\mathbb{N}$ and $k\in\{1,-1\}$.
\end{assumption}
By letting ties favor trust we obtain the weak inequality. We do this to facilitate a clean analysis. It should be noted however, that event in which equality holds has probability zero because the payoffs are continuous random variables.
We will often be interested in the case when both agents take the same action. As such we define $A(t)$ for $t\in \mathbb{N}$ without subscript as:
\begin{equation*}
    A(t) :=
    \begin{cases}
    T \quad &\text{if } A_{-1}(t) = A_{1}(t) = T \\
    D \quad &\text{if } A_{-1}(t) = A_{1}(t) = D \\
    0 &\text{else.}
    \end{cases}
\end{equation*}

When modeling agents that learn in the context of game theory, there is a variety of learning mechanisms which may be implemented. For example the agents could implement a basic form of reinforcement learning which would be an adjustment to our Assumption~\ref{As:Myopic}, combining belief and action on belief into one process. Alternatively, agents could make repeated application of Bayes' rule given the interaction they observe, or use stochastic approximation algorithm~\cite{Robbins1951} to update their belief on the probability that a random opponent will trust. The agents in our model use the exponential moving average~\cite{Brown1963} of their experiences. This is a straightforward, but powerful rule of thumb applied in signal processing. We introduce $\bar{\alpha}:=1-\alpha$ for improved legibility of future equations.  
\begin{assumption}[Belief updating]\label{As:ExpSmooth}
    The agents that are selected to play a game in round $t$ update their belief based on the outcome of that game using exponential smoothing with learning rate $\alpha\in(0,1)$:
\begin{equation*}
    x_{g(k,t)}(t+1):= \bar{\alpha}x_{g(k,t)}(t) +\alpha \mathbbm{1}{\{ A_{-k}(t) = T \}},
\end{equation*}
for $k\in\{1,-1\}$ and $t\in\mathbb{N}$. All other agents $i \in \{1,\ldots,N\}\setminus \{I(t),J(t)\}$ retain their most recent belief:
\begin{equation*}
    x_{i}(t+1):=x_i(t), \quad t\in\mathbb{N}.
\end{equation*}
\end{assumption}
We make this assumption mainly in order to facilitate a clean analysis. Furthermore, we believe this belief updating rule is simple enough for it not to be overly unrealistic to assume that people might learn in a similar fashion.

We note the work of Sato and collaborators~\cite{Sato2002,Sato2003,Sato2005} which also features updates according to the exponential moving average. An important difference between that work and ours is that we consider the discrete and random dynamics while they make use of a separation of time scales which forces actual dynamics to resemble the expectation. This last element is something, we shall see, emphatically not present in our model.

The learning rate $\alpha$ is the weight of an agents most recent observation and may thus be interpreted as the agent's memory. A greater weight to the most recent observation means that the earlier observations weigh less and thus are more quickly forgotten.

To an outsider who has not observed $\Pi_{k}(t)$ for $k\in\{1,-1\}$ during round $t\in\mathbb{N}$, the action taken looks random. Specifically, if the outsider is privy to the distributions of $U_k,Y_k,V_k$ and $W_k$, as well as the belief $x_k(t)$ then the probability that agent $k\in\{1,-1\}$ plays trust in round $t\in\mathbb{N}$ is defined as $p_k(t)$:
\begin{equation}
    p_k(t) := \mathbb{P}(u^T_{k}(t)\geq u^D_{k}(t)).\label{eq:imp_f}
\end{equation}
To clarify, for the agent $k\in \{-1,1\}$ who knows the values in $\Pi_k(t)$ and their own belief $x_k(t)$, the truth of the inequality $u^T_{k}(t)\geq u^D_{k}(t)$ is not random. For an outsider who is aware of the belief $x_k(t)$ and the distributions but not the realizations of the payoffs, the agent's actions seem random and follow the probability defined in (\ref{eq:imp_f}).
The behavior defined in (\ref{eq:imp_f}) may be restated for $k\in\{1,-1\}$ and $t\in\mathbb{N}$:
\begin{equation*}
    p_k(t)= \mathbb{P}(Z_k(t)\leq x_k(t)),
\end{equation*}
for which we define $Z_k(t)$ as the random variable:
\begin{equation}
    Z_k(t):=\frac{Y_k(t)-V_k(t)}{U_k(t)+Y_k(t)-V_k(t)-W_k(t)},
\end{equation}
for $k\in \{-1,1\}, t\in\mathbb{N}.$ We acknowledge the freedom we have in defining a cdf for $Z_k(t)$ by allowing correlation between its constituent random variables. Using this freedom we continue the discussion and analysis using $Z_k(t)$, and defining its cdf:
\begin{equation}
    F(x):=\mathbb{P}(Z_k(t)\leq x), \quad x\in \mathbb{R},
\end{equation}
which then also defines agent behavior $p_k(t)=F(x_k(t))$ for $k\in \{1,-1\}$ and $t\in\mathbb{N}$. We make the following assumption on $F$, which is essentially of a technical nature.
\begin{assumption}[Assumptions on $F$]\label{As:F}
    We consider only $F(\cdot)$ for which it is true that $F$ is a cumulative distribution function (non-decreasing and right-continuous) with $F(0)=0$, $F(1)=1$ and $F(x)\in (0,1)$ for all $x\in (0,1)$.
\end{assumption}
This assumption implies that if an agent believes that their opponent will trust (or doubt) with probability 1, that they too will trust (or doubt) with probability 1. Furthermore, that $F(\cdot)$ is a cdf and therefore non-decreasing, means that an agent with higher belief is at least as likely to trust than an agent with lower belief. Because $U_k(t),Y_k(t),V_k(t)$ and $W_k(t)$ are respectively iid between rounds $t\in\mathbb{N}$ as well as players $k\in\{-1,1\}$, the cdf $F(\cdot)$ does not change from round to round or from player to player. 

\begin{figure*}[htb]
    \centering
    \includegraphics[width = 0.9\textwidth]{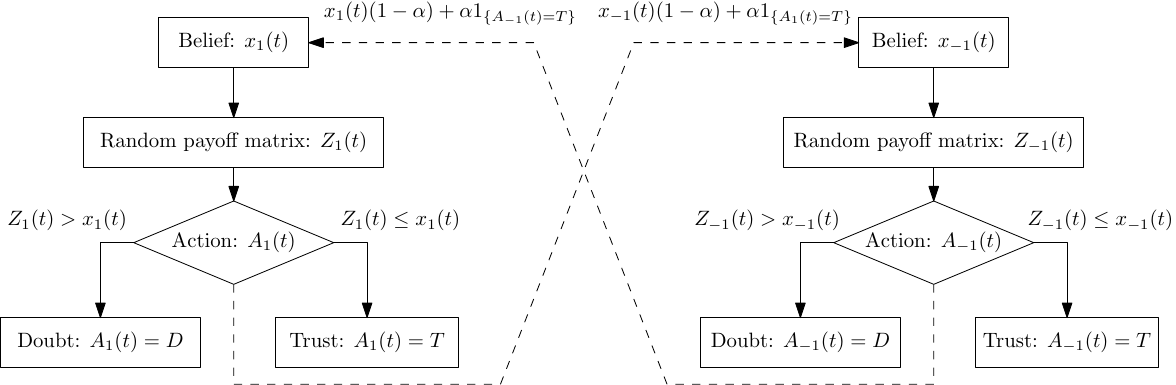}
    \caption{A flow chart of the model. This shows the process which the pair of randomly selected agents go through for one round. In each round two agents are drawn from the population at random to play the stochastic coordination game once against each other. Each of these agents observes an independent draw of the random payoff matrix and compares this with their belief. If their expected reward for doubting is greater than that for trust, then they doubt, and otherwise they trust. Then each of these agents adjusts its belief based on the action taken by its respective opponent and returns to the pool of possible agents to be selected in the next round.}
    \label{fig:flowchart}
\end{figure*}

The model we have described is highlighted by the flow chart in Figure~\ref{fig:flowchart}. This shows the process which the two randomly chosen agents follow in one given round. We are interested in the evolution of the belief vector $\belief{t}$ for $t\in\mathbb{N}$. Specifically we ask whether there is convergence of beliefs to an equilibrium state.

\section{Asymptotic analysis}
\label{Asym_as}
In this section we analyze the long-term dynamics of the process in the limit as time $t\to\infty$. We first show that it is possible for the $N$ agents to absorb (never exit an $\epsilon$-ball around) at $\bm{x}=\bm{0}$ and $\bm{x}=\bm{1}$ where agent behavior converges to always-doubt and always-trust respectively. Subsequently we show that the process may end up in these corners from the interior of the state space. Finally we combine these two sub-results in a Borel-Cantelli argument which proves that the process will converge to one of these corners in the long run with probability one. On an intuitive level this is explainable by the fact that in a coordination game, the agents will `try' to coordinate their behavior. Thus under learning and myopic rationality it is rational for a population to always trust or always doubt as this guarantees 100 percent coordination.

We state the main theorem, subsequently we will state and prove the lemmas required in the proof of this main theorem. We end this section by providing the proof of Theorem~\ref{thm:n_abs}.

In the statement of the main theorem, we use Lagrange notation to define $F^{(n)}(x)$ as the $n$-th derivative of $F$ at $x.$
\begin{theorem}[Absorption in a corner is guaranteed]\label{thm:n_abs}
If $\alpha>\epsilon$, and $F^{(n_1)}(0),F^{(n_2)}(1)\in [0,\infty)$ for some $n_1,n_2<\infty$ then 
\begin{multline}
    \mathbb{P}\left(\exists {t_0} : \{\bm{x}(t)\in [0,\epsilon]^N, \forall t>{t_0}\} \right.\\
    \cup \left. \{\bm{x}(t)\in [1-\epsilon,1]^N,\forall t>{t_0}\}\right) = 1,
\end{multline}
which may be written as
\begin{equation*}
    \mathbb{P}\left(\exists {t_0} :\bm{x}(t)\in [0,\epsilon]^N\cup [1-\epsilon,1]^N,\forall t>{t_0}\right) = 1.
\end{equation*}
\end{theorem}
The conditions on $F$ in this theorem are mild. All that we require of $F$ is that there exist finite integers $n_1$ and $n_2$ such that the $n_1$-th derivative of $F$ at $x=0$ exists and is finite and that the $n_2$-th derivative of $F$ at $x=1$ exists and is finite.
\subsection{Absorption is possible}
The first lemma we will need pertains to the possibility of a population converging in an $\epsilon$-ball around zero:
\begin{lemma}[Absorption at zero for $N$ agents]\label{lem:stay0N}
    Let $\alpha\in (0,1)$, $\epsilon\in (0,\alpha)$, and suppose $\bm{x}(t_0)\in [0,\epsilon]^N$ for some $t_0\in \mathbb{N}$, then
    \begin{equation}
    \mathbb{P}(\bm{x}(t)\in [0,\epsilon]^N, \forall t\geq t_0 \mid \bm{x}(t_0)\in [0,\epsilon]^N)
    >0.
\end{equation}
\end{lemma}
In most of the probabilities that we write there is a condition on the state of the system at $t_0$. In order to fit equations into the available space as well as for general legibility we define the following notation:
\begin{equation}
    \mathbb{P}_{[a,b]^N}(\cdot):= \mathbb{P}(\cdot \,| \,x(t_0)\in [a,b]^N).
\end{equation}

\begin{proof}
In the new notation, we will show that 
\begin{equation}
    \mathbb{P}_{[0,\epsilon]^N}(\bm{x}(t)\in [0,\epsilon]^N, \forall t\geq t_0)>0.
\end{equation}
    We prove this by induction. As base case for our induction we prove that absorption is possible for $N=2$:
    \begin{equation}
       q:= \mathbb{P}_{[0,\epsilon]^2}(\bm{x}(t)\in [0,\epsilon]^2, \forall t\geq t_0)>0. 
    \end{equation}

    When $\bm{x}(t)\in[0,\epsilon]^2$, then even one agent playing $T$ in some round $t_1\in\mathbb{N},t_1\geq t_0$ implies that $\bm{x}(t_1+1)\notin (0,\epsilon)^2$ because $x \bar{\alpha}+\alpha>\alpha>\epsilon$ for all $x\in (0,1)$ and $\alpha \in(0,1)$ and any $\epsilon\in (0,\alpha)$. Thus the probability of remaining in $[0,\epsilon]^2$ for all $t= t_0,t_0+1,\ldots,t_0+n$ is the probability of both agents playing doubt in all rounds from $t_0$ until (and including) round $t_0+n$:
    \begin{multline}
        \mathbb{P}_{[0,\epsilon]^2}(\bm{x}(t)\in (0,\epsilon)^2, t=t_0,\ldots,t_0+n)\\
        =\mathbb{P}_{[0,\epsilon]^2}(A(t)=D, t=t_0,\ldots,t_0+n).
    \end{multline}
    Note that we are interested in the limit of the above expression as $n\to\infty$ because the agents are required to play $D$ for all future rounds.
    The probability that agent $1$ plays $D$ in round $t$ is given by $1-F(x_1(t))$. For legibility we define $\bar{F}(\cdot):= 1-F(\cdot).$ Because the agents' actions during any round are independent, we have the above probability restated as
    \begin{multline}\label{eq:AlwaysD_new}
        \mathbb{P}_{[0,\epsilon]^2}(A(t)=D, t=t_0,\ldots,t_0+n) \\
        =\prod_{m=0}^n \bar{F}( \bar{\alpha}^m x_1(t_0))\bar{F}( \bar{\alpha}^m x_2(t_0)).
    \end{multline}
    In order to simplify expression we define
    \begin{equation}
        D_n:= \prod_{m=0}^n \bar{F}( \bar{\alpha}^m x_1(t_0))\bar{F}( \bar{\alpha}^m x_2(t_0)).
    \end{equation}
    Let $z_0:=\max \{x_1(t_0),x_2(t_0)\}$ and subsequently $z_m :=  \bar{\alpha}^m z_0$ $\forall m\in \mathbb{N}$. We bound (\ref{eq:AlwaysD_new}) by 
    \begin{equation*}
        D_n \geq \prod_{m=0}^n \bar{F}(z_m)^2
    \end{equation*}
    because $z_m\geq x_i(t_0) \bar{\alpha}^m$ for all $m\in\mathbb{N}$ and because $F(x)\geq F(y)$ as long as $x\geq y$ by virtue of being a distribution function.
    We take the logarithm of both sides as well as the limit as $n\to\infty$,
    \begin{equation*}
       \lim_{n\to\infty} \log\left(D_n\right)\geq \lim_{n\to\infty}\log\left( \prod_{m=0}^n (\bar{F}(z_m))^2\right),
    \end{equation*}
    and change the logarithm of a product on the right to a sum of logarithms:
    \begin{equation}\label{eq:logsum}
         \lim_{n\to\infty} \log\left(D_n\right)\geq \lim_{n\to\infty} 2\sum_{m=0}^n\log\left (\bar{F}(z_m)\right)
    \end{equation}
    We intend to bound the right hand side of (\ref{eq:logsum}). To do this observe first that $e^{-x}\geq 1-x$. This may be checked by evaluating the tangent of $e^{-x}$ at $x=0$ and noting that $e^{-x}$ is convex and so remains above this tangent line. We rearrange to obtain:
    \begin{equation*}
        1-e^{-x}\leq x, \quad \forall x\in \mathbb{R}.
    \end{equation*}
    We substitute $x=\log(y)$ to get
    \begin{equation*}
        1-e^{-\log(y)} \leq \log(y),
    \end{equation*}
    which simplifies to 
    \begin{equation}\label{eq:log_bound}
        1-\frac{1}{y} \leq \log(y), \quad \forall y\in \mathbb{R}_{>0}.
    \end{equation}
     We use the above inequality with $y=\bar{F}(z_m)$ to bound each term in the sum from below by 
     \begin{equation*}
         \log(\bar{F}(z_m))\geq 1-\frac{1}{\bar{F}(z_m)} = \frac{F(z_m)}{F(z_m)-1}.
     \end{equation*}
    Implanting this into the sum in (\ref{eq:logsum}) gives,
    \begin{equation}\label{eq:genBnd}
         \lim_{n\to\infty} \log\left(D_n\right)\geq \lim_{n\to\infty} 2\sum_{m=0}^n\frac{F(z_m)}{F(z_m)-1}.
    \end{equation}
    In order to invoke Abel's convergence test (see for example \cite{Whittaker1920}) we note that the sum in (\ref{eq:genBnd}) is the product of the sequences $\{F(z_m)\}$ and $\{1/(F(z_m)-1)\}$. The second of these is bounded from above by $-1$ and is monotone increasing in $m$. This is because $F(z_m)$ is monotone decreasing in $m$ and bounded by $0$. Therefore, if $\sum_{m=0}^\infty F(z_m)$ converges, then the right hand side of (\ref{eq:genBnd}) also converges.
    
    To prove convergence of $\sum_{m=0}^\infty F(z_m)$  we use the ratio test. We will show that
    \begin{equation}\label{eq:req_con_F}
        \lim_{m\to \infty} \left|\frac{F(z_{m+1})}{F(z_m)}\right|=\lim_{m\to\infty}\left|\frac{F( \bar{\alpha}^{m+1} z_0)}{F( \bar{\alpha}^m z_0)}\right|<1.
    \end{equation}
    The inequality is a result of an (possibly repeated) application of L'H\^{o}pital's rule.
    First note that 
    \begin{equation*}
        \lim_{m\to\infty}F(z_m)=\lim_{m\to\infty}F(z_{m+1})=0,
    \end{equation*}
    and so the limit of the quotient in (\ref{eq:req_con_F}) yields an indeterminate form $0/0$. We apply L'H\^{o}pital's rule:
  \begin{equation*}
         \lim_{m\to \infty} \left|\frac{F(z_{m+1})}{F(z_m)}\right| = \bar{\alpha}\frac{\lim_{m\to \infty}F'( \bar{\alpha}^{m+1}z_0)}{\lim_{m\to \infty}F'( \bar{\alpha}^{m}z_0)}.
    \end{equation*}
    \begin{itemize}
        \item Either at this point we have that $\sum_{k}F(z_m)$ converges if $F'(0)=c\in (0,\infty)$, or
        \item we apply L'H\^{o}pital's rule $n$ times until $F^{(n)}(0) =c\in (0,\infty)$ at which point the limit is $ \bar{\alpha}^n<1$.
    \end{itemize}
    Thus all conditions for Abel's convergence test are satisfied and we have shown convergence of the right hand side of (\ref{eq:genBnd}). In particular this will converge to a negative number because all the terms are negative as a result of the denominator. By taking the exponential of both sides of (\ref{eq:genBnd}) we get the probability of both agents playing doubt indefinitely on the left. On the right hand side we get a positive number. Considering that the probability of convergence is equal to the probability of both agents playing doubt indefinitely we have shown that this positive:
    \begin{equation*}
         \mathbb{P}_{[0,\epsilon]^2}(\bm{x}(t)\in [0,\epsilon]^2, \forall t\geq t_0)= \lim_{n\to\infty}D_n>0.
    \end{equation*}
This proves our base case: $q>0.$ Our induction hypothesis thus states that $N$ players can converge. As such we denote by $q_N$ the probability of $N$ players converging in an $\epsilon$-ball at zero:
\begin{equation}
    q_N:=\mathbb{P}_{[0,\epsilon]^N}(\bm{x}(t) \in [0,\epsilon]^N, \forall t\geq t_0 ) >0.
\end{equation}
Now we intend to show that 
    \begin{equation}
        \mathbb{P}_{[0,\epsilon]^{N+1}}(\bm{x}(t)\in [0,\epsilon]^{N+1}, \forall t\geq t_0)>0.
    \end{equation}
We identify (arbitrarily) the first agent and collect all the rounds in which agent $i=1$ is selected to play in $M:$
\begin{equation*}
    M:= \big\{t\geq t_0 : 1 \in \{I(t),J(t)\}\big\}.
\end{equation*}
We note that as before, if at least one agent plays trust in any round $t$ then $\bm{x}(t+1)\notin [0,\epsilon]^{N+1}$. Thus we have the relationship:
\begin{multline}
    \mathbb{P}_{[0,\epsilon]^{N+1}}(A(t) = D, \forall t\geq t_0)\\
    = \mathbb{P}_{[0,\epsilon]^{N+1}}(\bm{x}(t)\in [0,\epsilon]^{N+1}, \forall t\geq t_0).
\end{multline}
We may write the probability of all agents playing $D$ persistently as a product of conditional probabilities as follows:
\begin{widetext}
    \begin{equation}
    \mathbb{P}_{[0,\epsilon]^{N+1}}(A(t) = D, \forall t\geq t_0) \\
    =\prod_{j=0}^\infty \mathbb{P}_{[0,\epsilon]^{N+1}}(A(t_0+j)=D\mid A_k(\tau) = D, \forall t_0\leq\tau < t_0+j).
\end{equation}
\end{widetext}
Because we are dealing with multiplication on the right hand side, the ordering of terms does not matter. We are thus free to collect all the terms involving agent 1 in one product ($a$) and the remaining terms in a different product ($b$):
\begin{multline}\label{eq:def_a_b}
    \mathbb{P}_{[0,\epsilon]^{N+1}}(A(t) = D, \forall t\geq t_0) \\
    = \underbrace{\prod_{t\in M} \mathbb{P}_{[0,\epsilon]^{N+1}}(A(t)=D\mid A(\tau) = D, \forall t_0\leq\tau < t)}_{=:a} \\
    \times \underbrace{\prod_{t\notin M }\mathbb{P}_{[0,\epsilon]^{N+1}}(A(t)=D\mid A(\tau) = D , \forall t_0\leq\tau < t)}_{=:b}.
\end{multline}

For the rounds $t\in M$ we introduce the notation $t_{i,l}$ to be the time of the $l$-th round in which agent 1 is chosen to play against agent $i$. Note that because the dynamics never end, agent one will be chosen to play with each other agent for an infinite number of rounds. Thus we can split $a$ further and in terms of agent beliefs:
\begin{equation*}
    a = \prod_{i=2}^{N+1}\prod_{l=1}^\infty \bar{F}(x_1(t_{i,l}))\bar{F}(x_i(t_{i,l})).
\end{equation*}
We can bound the beliefs of the agents involved because we know that at least they have played against each other $l$ times, and always played doubt (by the conditioning). So we set $\tilde{x}_l:=\epsilon  \bar{\alpha}^{l-1}$ for $l = 1,2,\ldots$ and note that $x_j(t_{i,l})\leq \tilde{x}_l$ for both $j=1,$ and $i$ as well as all $l = 1,2,\ldots.$

We can bound the probability in the product by:
\begin{equation*}
    a\geq \prod_{i=2}^{N+1}\prod_{l=1}^\infty \bar{F}(\tilde{x}_l)^2 = \prod_{l=1}^\infty (\bar{F}(\tilde{x}_l))^{2N}.
\end{equation*}
We take the logarithm on both sides:
\begin{equation*}
    \log(a) \geq 2N\sum_{l=1}^\infty  \log (\bar{F}(\tilde{x}_l)).
\end{equation*}
From here we can repeat the steps followed in the proof of our base case from (\ref{eq:logsum}) to conclude that the sum converges. This means that $a>0$ as long as $N < \infty$.

We proceed to show that $b>0$ (which is defined in (\ref{eq:def_a_b})). We note that $b>q_N$, the probability of $N$ agents absorbing at zero. To see that this is true consider the last $N$ agents and some order of games for them to play against each other. By the induction hypothesis this group of $N$ players have positive probability of always playing doubt. But this group and these games are interspersed with matches between some agent in the group of $N$ and the first agent. We have conditioned on both agents playing doubt in all games until the current game which includes those involving the first agent. This implies that the beliefs of agents who were chosen for games against the first agent have a lower belief after this game and play doubt at an even higher probability in their next match than in the original set of games in which there was already a probability $q>0$ of all agents always doubting. 

We have thus shown that $a>0$ and $b>0$ and therefore conclude that also $a\cdot b>0$.
\end{proof}

We now state a similar lemma for absorption of $N$ agents around one.
\begin{lemma}[Absorption at one for $N$ agents]\label{lem:stay1N}
    Let $\alpha\in (0,1)$, $\epsilon\in (0,\alpha)$, and suppose $\bm{x}(t_0)\in [1-\epsilon,1]^N$ for some $t_0\in \mathbb{N}$, then
    \begin{multline}
    \mathbb{P}(\bm{x}(t)\in [1-\epsilon,1]^N, \forall t\geq t_0 \mid \bm{x}(t_0)\in [1-\epsilon,1]^N)\\
    = \mathbb{P}_{[1-\epsilon,1]^N}(\bm{x}(t)\in [1-\epsilon,1]^N, \forall t\geq t_0)>0.
\end{multline}
\end{lemma}
The proof is similar to the proof of Lemma~\ref{lem:stay0N}. The difference is that instead of playing doubt the agents are required to play trust indefinitely. This happens at probability $F(x)$ rather than $\bar{F}(x).$ Furthermore the agent belief at the start of the $l$-th round after $t_0$ is $1-(1-x(t_0)) \bar{\alpha}^{l-1}$ rather than $x(t_0) \bar{\alpha}^l$. 

Intuitively our first two lemmas imply that a population that is within an $\epsilon$-ball around zero or one, can remain there. This means a population believing that everyone is 100\% (or almost 100\%) trustworthy or untrustworthy, can retain that belief forever.

\subsection{Reaching the corners is possible}
Our next result proves sufficient conditions for the population to reach the $\epsilon$-ball around zero with positive probability. For legibility we define:
\begin{equation}
    I_N:=[0,1]^N\setminus\left([0,\epsilon]^N \cup [1-\epsilon,1]^N\right),
\end{equation}
which we call the \textit{interior}.
\begin{lemma}[Population of $N$ agents reaches zero with positive probability.]\label{lem:n_reach0}
    Let $\alpha \in (0,1)$ and $\bm{x}(t_0)\in I_N$ for some $t_0\in\mathbb{N}$ and $\epsilon\in (0,\alpha)$, then 
    \begin{equation*}
        \mathbb{P}_{I_N}(\exists t_1\in \mathbb{N}:\bm{x}(t_1)\in [0,\epsilon]^N, t_1>t_0)>0. 
    \end{equation*}
\end{lemma}
\begin{proof}
We construct a path from $I_N$ which depends on agents always playing doubt against one another. 
    
We split this into two cases, in the first we show that there is a path from $\bm{x}(t_0)\in (\epsilon,1-\epsilon)^N$ to $(0,\epsilon)^N$ of positive probability. In the second we show that there is a path from $\bm{x}(t_0)\in I_N$ with some number $0<h<N$ of agents with belief $x(t_0)\geq1-\epsilon$ to $(\epsilon,1-\epsilon)^N$ of positive probability and therefore also a path to $(0,\epsilon)^N$.

\textbf{Case 1:} Suppose all agents have belief $x_i(t_0)<1-\epsilon$. From now until round $m$ agent $2k-1$ is matched to play against agent $2k$ where $k =r\mod{N/2}$~\footnote{This proof handles the case where $N$ is even. To extend the result to all odd $N\geq3$, a simple extension is possible: The first agent plays a game with last agent at the start of each iteration step and then joins the pool of $N-1$ agents which proceeds as in the case we handle.} where $r=t-t_0$. The probability of each round of this matching is given by $1/(N(N-1))$. The probability of this pattern of matching for $m$ rounds is then given by
    \begin{equation*}
        p_m :=\prod_{r=1}^m \frac{1}{N}\frac{1}{N-1} = \frac{1}{N^m(N-1)^m}.
    \end{equation*}
    Note that each pair of agents ($2k-1$ and $2k$) is independent of the other agents for these $m$ rounds, in which they each play $2m/N$ games. Let $t_{i,l}$ for $l = 1,2,\ldots,2m/N$ index the time of the round in which agent $i$ plays their $l$-th game (after $t_0$), then supposing all agents play $D$ in all rounds until round $m$, then agent $i$'s belief follows:
    \begin{equation*}
        x_i(t_{i,l}) =  \bar{\alpha}^{l}x_i(t_0), \quad \text{for all }l = 1,2,\ldots, 2m/N.
    \end{equation*}
    Let $\kappa_i\in\mathbb{N}$ be the minimum number of games agent $i$ has to play (both players always playing doubt $D$) for their belief to be distance $\epsilon$ from $0$. Specifically $\kappa_i$ is the least value that satisfies:
    \begin{equation}\label{eq:k_def}
          \bar{\alpha}^{\kappa_i}x_i(t_0)<\epsilon.
    \end{equation}
    By dividing through by $x_i(t_0)>0,$ and taking the logarithm we see that 
    \begin{equation}\label{eq:k_finite}
    \kappa_i = \bigg\lceil \frac{\log (\epsilon/x_i(t_0))}{\log  \bar{\alpha}}\bigg\rceil.
\end{equation}
This is finite for all $x_i(t_0)\leq1-\epsilon, \epsilon>0$ and $\alpha<1$ and is maximized at $x_i(t_0)=1-\epsilon.$

For $\bm{x}(t)\in (0,\epsilon)^N$ we need $x_i(t)<\epsilon$ to be true for all agents $i = 1,2,\ldots, N$. We define $m$ to be the least value $m$ such that each agent has played enough games to be less than distance $\epsilon$ from zero:
\begin{equation*}
    m:=\min\{m \in \mathbb{N}: 2m/N \geq k_i, \forall i =1,2,\ldots, N\}.
\end{equation*}
By rearrangement we have 
\begin{equation}
    m= \max_{i=1,2,\ldots, N}\left\{\big\lceil  \frac{N\cdot\kappa_i}{2}\big\rceil \right\},
\end{equation}
which is finite because $\kappa_i$ is finite. This gives us our first intermediate result: The probability of the matching we have created is thus strictly greater than zero. We call this $p_m>0.$

We now turn to the other requirement of this path to $(0,\epsilon)^N$ which is that all agents play $D$ in each of their $2m/N <\infty$ games. As noted before, each pair of agents interacts exclusively with one another and so we focus on one such pair and the rounds in which they play. For pair $k = 1, 2, \ldots, N/2$ denote the probability that there exists a round $t_k$ for which $x_{2k-1}(t_k),x_{2k}(t_k)<\epsilon$:
\begin{multline}
    \mathbb{P}_{I_N}(\exists t_k \in \mathbb{N} : x_{2k-1}(t_k),x_{2k}(t_k)<\epsilon) \\\geq \prod_{l=0}^{2m/N} \bar{F}( \bar{\alpha}^lx_{2k-1}(t_0))\bar{F}( \bar{\alpha}^lx_{2k}(t_0)).
\end{multline}
Let $\tilde{x}_{k,0}:=\max\{x_{2k-1},x_{2k}\}$ and subsequently $\tilde{x}_{k,l} =  \bar{\alpha}^l\tilde{x}_{k,0}$. We use this to bound the beliefs of the $k$-th pair in their $l$-th interaction. Then we can bound the right hand side:
\begin{multline}
     \mathbb{P}_{I_N}(\exists t_k \in \mathbb{N} : x_{2k-1}(t_k),x_{2k}(t_k)<\epsilon)\\
     \geq \prod_{l=0}^{2m/N} (\bar{F}(\tilde{x}_{k,l}))^2,
\end{multline}
because $\tilde{x}_{k,l}\geq x_{2k-1}(t_0) \bar{\alpha}^l, x_{2k}(t_0) \bar{\alpha}^l$ for all $l\in\mathbb{N}$ and because $F(x)\geq F(y)$ as long as $x\geq y$ by virtue of being a distribution function. The terms in the product are all strictly greater than 0 and because it is a finite product we know that its result is also strictly greater than 0 for all $k = 1, 2,\ldots, N/2$. Finally we have $N/2$ of these pairs and so the probability of reaching the corner is the probability of the matching times the probability of the individual games proceeding as described:
\begin{multline}
    \mathbb{P}_{I_N}(\exists t_1\in \mathbb{N}:\bm{x}(t_1)\in [0,\epsilon]^N, t_1>t_0)\\
    > p_m \cdot \mathbb{P}_{I_N}(\exists t_k \in \mathbb{N} : x_{2k-1}(t_k),x_{2k}(t_k)<\epsilon)^{N/2} >0.
\end{multline}
This proves the statement in case 1 where all the agent beliefs were $x_i(t_0)\leq 1-\epsilon$ for all $i = 1,2,\ldots, N$. 

\textbf{Case 2:} Alternatively there are agents who have belief $x_i(t_0)>1-\epsilon$. As worst case scenario, suppose $x_1(t_0) = 1-\epsilon$, and $x_j(t_0)=1$, for $j=2,\ldots, N$. Here we have chosen the first agent to be the agent with belief $x(t_0)\leq 1-\epsilon$ without loss of generality as their naming convention plays no role. We now construct a finite path of positive probability to get from this state to the state we assumed at the start of the proof ($x_i(t_0)<1-\epsilon$ for all $i=1,2,\ldots, N$).

For $k =1,2,\ldots, N-1$, we specify the games between $I$ and $J$ as follows:
\begin{multline*}
    (I(t),J(t))^*_{t=t_0+1, \ldots, t_0+2(N-1)}\\
    =\begin{cases}
        I(t_0+2k-1)=I(t_0+2k) \in \{1,\ldots,k\} \\
        J(t_0+2k-1)=J(t_0+2k)  = k+1.
    \end{cases}
\end{multline*}
The probability of such a sequence of matches is:
\begin{multline}
   p_m' :=\mathbb{P}\left((I(t),J(t))^*\right)\\
   =\prod_{k=1}^{N-1} \left(\frac{k}{N}\right)\left(\frac{1}{N}\right)\left(\frac{1}{N-1}\right)^2 >0.
\end{multline}
The above is a finite product of positive numbers and so is greater than zero. The general form of this sequence is that some agent from the set $\{1,\ldots, k\}$ plays against the first agent that isn't in the set twice. In the first of these two games, player $I$ plays doubt in both rounds, while player $J$ plays trust in the first round, and doubt in the second round. In this way, after the two games, both agents have $x<1-\epsilon$. 

To see that this is the case consider agent $I$ whose belief starts at $x(t_0)\leq 1-\epsilon.$ After playing against $T$ their belief updates to $x(t_0+1)\leq 1-\epsilon  \bar{\alpha}<1.$ Subsequently they play against $D$ and so in the following round they hold belief $x(t_0+2)\leq 1-\alpha -\epsilon  \bar{\alpha}^2 < 1-\alpha < 1-\epsilon.$ Now we consider the agent $J$ whose belief starts at $x(t_0) \leq 1$. After playing against $D$ twice their belief is bounded by $x(t_0+2)\leq  \bar{\alpha}^2<1-\alpha<1-\epsilon.$

For one such pair of games define $B$ the event of the sequence of actions (starting in round $2k+1$):

\begin{equation*}
\begin{split}
    B:=&\{A_1(t+2k-1)=D\}\\
    &\cap \{A_{-1}(t+2k-1)=T\}\cap \{A(t+2k)=D \}.
\end{split}
\end{equation*}

We call $p_{2g}$, the probability of both players acting according to the event $B$, and we note that it is bounded:
\begin{equation}\label{eq:twoG}
    p_{2g}:=\mathbb{P}_{I_N}(B)
    \geq\bar{F}(1-\epsilon)\cdot 1 \cdot\bar{F}(1-\epsilon  \bar{\alpha}) \cdot \bar{F} \bar{\alpha}.
\end{equation}
The right hand side of (\ref{eq:twoG}) is strictly greater than zero because by Assumption~\ref{As:F} $F$ is only zero at zero, and so we have product of 4 numbers, all greater than zero. Thus $p_{2g}>0.$

The probability then of getting into case~1 is given by the probability of the matching $(I,J)^*$ multiplied by the probability of all $(N-1)$ pairs of games going as planned which are independent and the probability is:
\begin{equation}
    p_m' \cdot p_{2g}^{N-1}>0.
\end{equation}
At time $t_0+2(N-1)+1$ all the agents' belief is $x<1-\epsilon.$ Proceeding as in case 1 we know that there is a path of positive probability to $(0,\epsilon)^N.$
\end{proof}
We now state a similar lemma for the probability of the population reaching the $\epsilon$-ball around 1 from the interior.
\begin{lemma}[Population of $N$ agents reaches one with positive probability.]\label{lem:n_reach1}
    Let $\alpha \in (0,1)$ and $\bm{x}(t_0)\in I_N$ for some $t_0\in\mathbb{N}$ and $\epsilon\in (0,\alpha)$, then 
    \begin{equation*}
        \mathbb{P}_{I_N}(\exists t_1\in \mathbb{N}:\bm{x}(t_1)\in [1-\epsilon,1]^N, t_1>t_0)>0. 
    \end{equation*}
\end{lemma}
The proof is similar to that of Lemma~\ref{lem:n_reach0}. The differences are akin to the differences between Lemma~\ref{lem:stay0N} and Lemma~\ref{lem:stay1N}. Additionally, instead of looking for a least $\kappa_i$ that satisfies (\ref{eq:k_def}) we look for a least $\kappa_i$ that satisfies $1-(1-x(t_0)) \bar{\alpha}^{\kappa_i}>1-\epsilon$. This can however be translated into (\ref{eq:k_def}). We now state the main analytical result of the paper. By the dynamics we describe, a population of any finite size $N$, is guaranteed to converge, in belief as well as behavior, to one of the corners $\bm{0}$ or $\bm{1}$ of the state space $[0,1]^N$.

Our second pair of lemmas tell us that a population can end up believing that the rest of the population is 100\% trustworthy (or untrustworthy). Thus a natural reinforcement of beliefs regarding trustworthiness can take place and result in complete trust among the population or a complete lack thereof.

\subsection{Proof of the main theorem}
We now present the proof of the main theorem of our paper which states that a population of agents will converge with probability one at either the always-trust or the always-doubt corner.

\begin{proof}
Our objective is to show that if $\alpha>\epsilon$, and $F^{(n_1)}(0),F^{(n_2)}(1)\in [0,\infty)$ for some $n_1,n_2<\infty$ then 
\begin{equation*}
    \mathbb{P}\left(\exists {t_0} :\bm{x}(t)\in [0,\epsilon]^N\cup [1-\epsilon,1]^N,\forall t>{t_0}\right) = 1.
\end{equation*}
By Lemmas~\ref{lem:n_reach0} and~\ref{lem:n_reach1} we know that the process may reach $A:=[0,\epsilon]^N\cup [1-\epsilon,1]^N$ from $[0,1]\setminus A$ in $s$ (a finite number of) rounds with positive probability.

By Lemma~\ref{lem:stay0N} and~\ref{lem:stay1N} we know that the belief vector, upon reaching $A$ has positive probability of being absorbed there. As such we define the probability of reaching and being absorbed in $A$ in $s+1$ rounds as:
\begin{equation*}
    p:=\mathbb{P}(\bm{x}(t)\in A,\forall t\geq t_0+s+1),
\end{equation*}
and note that $p>0$. Define $\tau_1$ as the time of the last entry into $A$:
\begin{equation*}
\tau_1 := \min\{\tau: \bm{x}(t)\in A, \forall t\geq \tau\} \in \mathbb{N} \cup \{\infty\}.
\end{equation*}
Thus after $s+1$ rounds (supposing we can read the future about possible absorption) one of three things will have occurred:
\begin{enumerate}
    \item The belief vector $\bm{x}(t_0+s+1)$ is absorbed in $A$, and so $t_0+s+1 \geq \tau_1$.
    \item The belief vector $\bm{x}(t_0+s+1)$ is not absorbed in $A$ (thus $t_0+s+1 < \tau_1$) and is somewhere in $[0,1]^N\setminus A$.
    \item The belief vector $\bm{x}(t_0+s+1)$ is in $A$ (thus $t_0+s+1 < \tau_1$) but has not been absorbed. 
\end{enumerate}
If (3) happens then at some point the belief will have to again be in $[0,1]^N\setminus A$ (with probability 1 as this is implied by not being absorbed). Once the process is again in $[0,1]^N\setminus A$ either directly in case of (2) and after some finite time in case (3), we reset the clock and note that after another $s+1$ rounds the process is either absorbed in $A$ (with probability $p$) or not (with probability $1-p$). We therefore know that the probability of the belief vector $\bm{x}$ not being absorbed in $A$ after $n\in \mathbb{N}$ such meta-experiments is given by $(1-p)^n$ and the probability of never being absorbed is:
\begin{equation*}
    \mathbb{P}(\nexists t_0: \bm{x}(t)\in A,\forall t\geq t_0) = \lim_{n\to \infty}(1-p)^n =0.
\end{equation*}
Thus the probability of the complement is $\mathbb{P}(\exists t_0: \bm{x}(t)\in A,\forall t\geq t_0) =1$.
\end{proof}
In this section we showed that it is certain that the process ends in one of the corners in the long run~\footnote{Note that it might be possible to prove this result by carefully checking whether the dynamics of this model satisfy the conditions in Norman~\cite{Norman1968}. We do not believe this to be a straightforward task.}. The natural next question is: With what probability does the system absorb in the always-trust corner? Define the probability of (eventual) absorption in the always-trust corner $\bm{x}\in [1-\epsilon,1]^N$ as:
\begin{equation*}
    p_T(\bm{x}(t_0)):=\mathbb{P}(\exists t_1: \bm{x}(t)\in [1-\epsilon,1]^N, \forall t\geq t_1). 
\end{equation*}
Similarly we define the probability of absorption in the always doubt corner:
\begin{equation*}
     p_D(\bm{x}(t_0)):=\mathbb{P}(\exists t_1: \bm{x}(t)\in [0,\epsilon]^N ,\forall t\geq t_1). 
\end{equation*}
Because absorption in one of the two corners is guaranteed by Theorem~\ref{thm:n_abs}, we have the following relationship for all $\bm{x}\in [0,1]^N$:
\begin{equation*}
    p_D(\bm{x}) = 1-p_T(\bm{x}).
\end{equation*}
By this result we are able to focus our investigation into the dichotomy of absorption at $\bm{1}$ or at $\bm{0}$, and not miss dynamics in which the process reaches a steady state in the interior of the state space. 
\begin{remark}
    The main result of this section, Theorem~\ref{thm:n_abs} also holds true for any finite connected population structure determining agent pairings. Absorption of the process $\bm{x}$ at $\bm{0}$ or $\bm{1}$ is guaranteed in the long-term.
\end{remark}
The proofs for the corresponding results of Lemmas~\ref{lem:stay0N}--\ref{lem:n_reach1}, would need to be adjusted by accounting for the restrictions on which agents may interact with one another. With some care on the order of agent pairings, as long as the network of agents in connected, it should be possible to construct a sequence of pairings which leads to reaching the $\epsilon$-balls at zero and one. The same holds true for absorption. 

\section{Illustration for two agents}
\label{Illustration}
In this section we illustrate the model by means of exploring the two agent case. Suppose that $N=2$, then we have two agents who are matched to play against one another in all rounds $t=1,2,\ldots$. Note that the agents assignment to $I$ or $J$ is merely a matter of notation. Thus we can assume that agent 1 is always assigned $k=1$ and agent 2 is assigned $k=-1$ and refer to them by their $k$ assignment for the remainder of this section.

\begin{figure}[b]
    \centering
    \includegraphics[width = 0.45\textwidth]{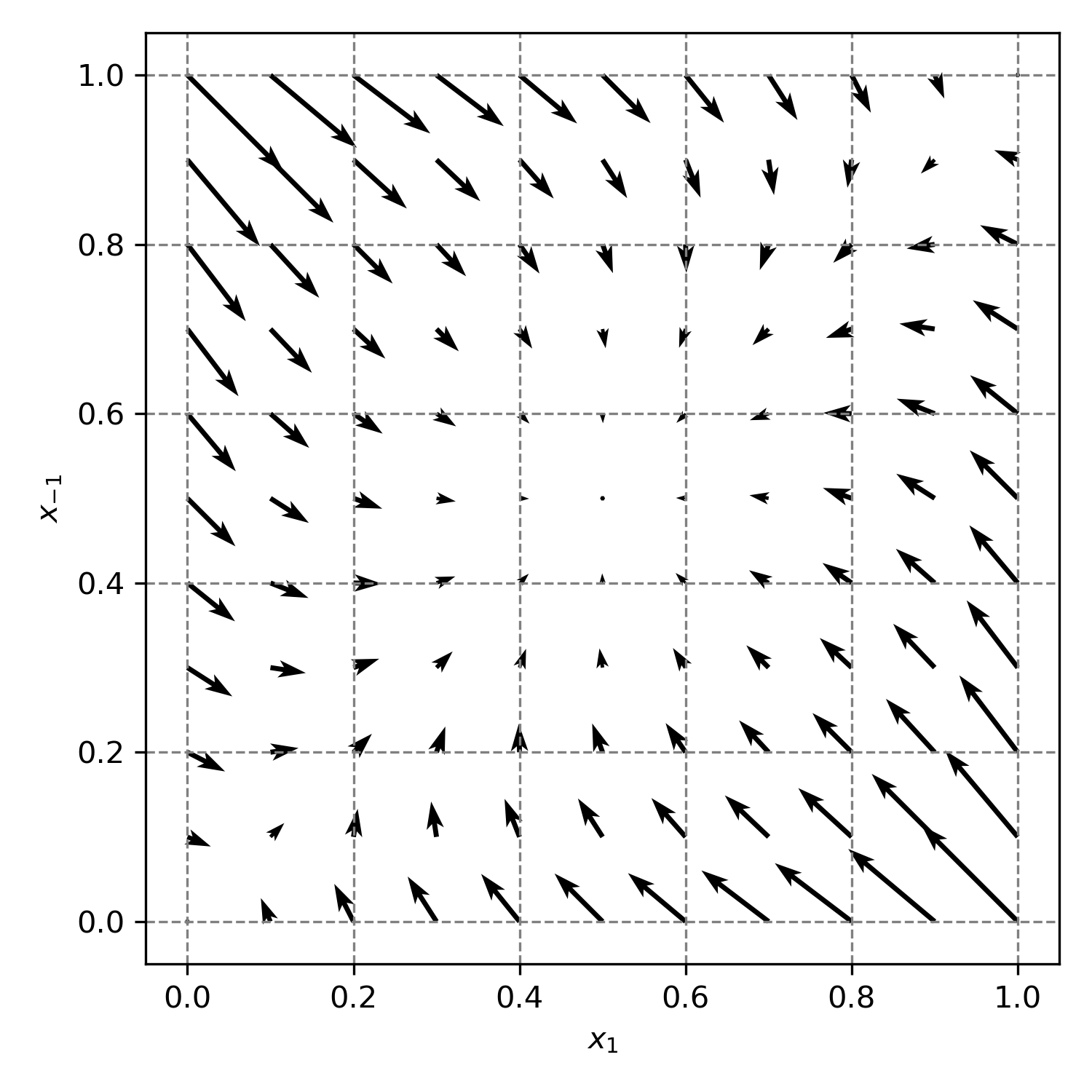}
    \caption{The expected value of $h(\bm{x})-\bm{x}$ for the function $F(x) = x+2x(x-1/2)(x-1).$ This figure highlights the at first counter intuitive nature of the result in Theorem~\ref{thm:n_abs}. The expectation of the dynamics would indicate that the process absorbs at $\bm{x}=(0.5,0.5)$, but we know by Theorem~\ref{thm:n_abs} that the process absorbs with probability one at one of the two corners $\bm{x}=(0,0)$ or $(1,1)$.}
    \label{fig:Vec_fields}
\end{figure}

We have the following stochastic dynamical system:
\begin{equation}
    \begin{bmatrix}
         x_1(t+1) \\
    x_{-1}(t+1)
    \end{bmatrix}= h(\bm{x}(t))+\xi_t, 
\end{equation}
where $\xi_t$ is defined as the vector of noise which accounts for the randomness:
\begin{equation}
    \xi_t:=
    \begin{bmatrix}
         \bar{\alpha}x_1(t) + \alpha \mathbbm{1}_{\{A_{-1}(t)=T\}}\\
     \bar{\alpha}x_{-1}(t) + \alpha \mathbbm{1}_{\{A_{1}(t)=T\}}
    \end{bmatrix},
\end{equation}
and we define $h(\cdot)$ to be the expected evolution of the system. We may calculate $h(\bm{x})$ explicitly in terms of the cdf $F:$
\begin{equation*}
    h(\bm{x}(t)) = 
    \begin{bmatrix}
         \bar{\alpha}x_1(t) + \alpha F(x_{-1}(t))\\
         \bar{\alpha}x_{-1}(t) +\alpha F(x_1(t))
    \end{bmatrix}.
\end{equation*}
We note that any points with $x_1 = F(x_{-1})$ and $x_{-1}=F(x_1)$ map to themselves in expectation. Furthermore, because $F(x)\geq F(y)$ whenever $x\geq y$ (because $F$ is a distribution function) the points that map to themselves in expectation must satisfy:
\begin{align}
    x_1 &= x_{-1},\\
    x_1 &= F(x_1).
\end{align}
For the function $F(x) = x+2x(x-1/2)(x-1)$, we plot the vector field of the expected change: $h(\bm{x})-\bm{x}$ in Figure~\ref{fig:Vec_fields}. Looking at the regions $(0,0.4)^2$ and $(0.6,1)^2$ in Figure~\ref{fig:Vec_fields}, we might guess that when $F(x) = x+2x(x-1/2)(x-1)$ convergence will be to the center of the domain: $\bm{x}=(0.5,0.5)$. However, by Theorem~\ref{thm:n_abs} we know that the process must converge in one of the two corners in the long run.

Heuristically this can be explained by the progression of agent beliefs around zero and one versus around $0.5$. There are paths by which the agents' beliefs can keep getting closer to zero and one respectively. Meanwhile, at $0.5$, should the agents' beliefs get arbitrarily close, their next belief will be roughly $\vert 0.5 \alpha\vert$ away from $0.5$.

In short; knowing the result of Theorem~\ref{thm:n_abs} and comparing this with Figure~\ref{fig:Vec_fields}, we realize that we should be careful not to assume that the asymptotic dynamics will be dictated by their expectation.

\section{Simulation}
\label{Simulation}
We analyze the effect of $F$ and $\alpha$ on $p_T$ and the time until first entry to $A$ by means of simulation. Because it requires substantially less simulated time steps to simulate the process with $N=2$ agents, we begin in this setting, and later include a numerical simulation to verify that the results are qualitatively similar to the case when $N>2$. 

\begin{figure}[h!]
    \centering
    
    \subfloat[$\alpha=0.01$]{\resizebox{0.38\textwidth}{!}{
\begin{tikzpicture}

\definecolor{darkgray176}{RGB}{176,176,176}
\definecolor{darkorange25512714}{RGB}{255,127,14}
\definecolor{forestgreen4416044}{RGB}{44,160,44}
\definecolor{lightgray204}{RGB}{204,204,204}
\definecolor{steelblue31119180}{RGB}{31,119,180}

\begin{axis}[
legend cell align={left},
legend style={
  fill opacity=0.8,
  draw opacity=1,
  text opacity=1,
  at={(0.5,0.5)},
  anchor=center,
  draw=lightgray204
},
tick align=outside,
tick pos=left,
unbounded coords=jump,
x grid style={darkgray176},
xlabel={\large{$t$}},
ticklabel style = {font = \large},
xmin=-373.95, xmax=7874.95,
xtick style={color=black},
y grid style={darkgray176},
ylabel={\large{$|x(t)-A|$}},
ymin=-0.0341688221710105, ymax=0.719745265591219,
ytick style={color=black},
ytick={-0.2,0,0.2,0.4,0.6,0.8},
yticklabels={\ensuremath{-}0.1,0.0,0.1,0.2,0.3,0.4}
]
\addplot [very thick, steelblue31119180]
table {%
1 0.659570581000001
5 0.663164602
10 0.666572968000001
30 0.672693742000001
50 0.669580815000001
100 0.622613421
250 0.342424993
500 0.103403329
750 0.0297955019999999
1000 0.00839902199999989
1500 0.000783004999999975
2000 0.000257360999999986
2500 0.000212596999999987
3000 0.000193435999999988
3500 0.000185459999999987
4000 0.000189162999999989
5000 0.000198209999999988
5500 0.000209010999999988
7500 0.000186385999999988
10000 nan
};
\addlegendentry{$r = 0.5$}
\addplot [semithick, steelblue31119180, opacity=0.5, forget plot]
table {%
1 0.649116229762505
5 0.652767301872132
10 0.656257703408744
30 0.662622219262357
50 0.659567383498575
100 0.612197759952041
250 0.334893057788676
500 0.100767819123867
750 0.0287833805288679
1000 0.00795030710880543
1500 0.00067142857786857
2000 0.000211645435829063
2500 0.000165990807322302
3000 0.000140912764626789
3500 0.00014648459051903
4000 0.000156374981829214
5000 0.000158365543683718
5500 0.000159639676866408
7500 0.000148024614496515
10000 nan
};
\addplot [semithick, steelblue31119180, opacity=0.5, forget plot]
table {%
1 0.670024932237497
5 0.673561902127869
10 0.676888232591257
30 0.682765264737645
50 0.679594246501426
100 0.63302908204796
250 0.349956928211324
500 0.106038838876133
750 0.0308076234711318
1000 0.00884773689119435
1500 0.000894581422131379
2000 0.00030307656417091
2500 0.000259203192677673
3000 0.000245959235373187
3500 0.000224435409480945
4000 0.000221951018170764
5000 0.000238054456316257
5500 0.000258382323133568
7500 0.000224747385503461
10000 nan
};
\addplot [very thick, dotted, darkorange25512714]
table {%
1 0.673851776
5 0.673900734999999
10 0.673955406
30 0.67506003
50 0.67419943
100 0.671233506999999
250 0.664051923999999
500 0.647958422
750 0.636744196
1000 0.630773511000001
1500 0.610937789999999
2000 0.590142117
2500 0.578051605999999
3000 0.557744890999999
3500 0.53735747
4000 0.523012214
5000 0.501566970999999
5500 0.490214824999999
7500 0.451604643999998
10000 nan
};
\addlegendentry{$r = 1.0$}
\addplot [semithick, darkorange25512714, opacity=0.5, forget plot]
table {%
1 0.663499522220714
5 0.663531185899315
10 0.663602494052429
30 0.66464361657979
50 0.663761013842673
100 0.660798745786364
250 0.653455334556737
500 0.637103931967174
750 0.625536965851429
1000 0.619369233531258
1500 0.59935593978913
2000 0.578352995024377
2500 0.565971031003422
3000 0.545240022265156
3500 0.524611036112559
4000 0.509957888593191
5000 0.488005595045376
5500 0.476619398435852
7500 0.43746728139228
10000 nan
};
\addplot [semithick, darkorange25512714, opacity=0.5, forget plot]
table {%
1 0.684204029779286
5 0.684270284100683
10 0.684308317947572
30 0.685476443420209
50 0.684637846157326
100 0.681668268213633
250 0.674648513443262
500 0.658812912032826
750 0.647951426148571
1000 0.642177788468744
1500 0.622519640210868
2000 0.601931238975623
2500 0.590132180996576
3000 0.570249759734843
3500 0.550103903887441
4000 0.536066539406808
5000 0.515128346954621
5500 0.503810251564146
7500 0.465742006607717
10000 nan
};
\addplot [very thick, dashed, forestgreen4416044]
table {%
1 0.659053947000001
5 0.658546804
10 0.657167201000001
30 0.648264344
50 0.62888533
100 0.539742464
250 0.215580518
500 0.023680883
750 0.00201414699999999
1000 0.000190619999999994
1500 0.000100000000000002
2000 0.000100000000000002
2500 0.000100000000000002
3000 0.000100000000000002
3500 0.000100000000000002
4000 0.000100000000000002
5000 0.000100000000000002
5500 0.000100000000000002
7500 0.000100000000000002
10000 nan
};
\addlegendentry{$r = 2.0$}
\addplot [semithick, forestgreen4416044, opacity=0.5, forget plot]
table {%
1 0.648725766747944
5 0.648235583944514
10 0.64687357811062
30 0.637833017545143
50 0.618167954941555
100 0.528538067472763
250 0.207427168498144
500 0.0220611139002947
750 0.00185103570261596
1000 0.00017778980125517
1500 0.000100000000000002
2000 0.000100000000000002
2500 0.000100000000000002
3000 0.000100000000000002
3500 0.000100000000000002
4000 0.000100000000000002
5000 0.000100000000000002
5500 0.000100000000000002
7500 0.000100000000000002
10000 nan
};
\addplot [semithick, forestgreen4416044, opacity=0.5, forget plot]
table {%
1 0.669382127252057
5 0.668858024055486
10 0.667460823889381
30 0.658695670454856
50 0.639602705058445
100 0.550946860527238
250 0.223733867501856
500 0.0253006520997053
750 0.00217725829738402
1000 0.000203450198744817
1500 0.000100000000000002
2000 0.000100000000000002
2500 0.000100000000000002
3000 0.000100000000000002
3500 0.000100000000000002
4000 0.000100000000000002
5000 0.000100000000000002
5500 0.000100000000000002
7500 0.000100000000000002
10000 nan
};
\end{axis}

\end{tikzpicture}\label{fig:pre_a1}}}\\
    \subfloat[$\alpha = 0.05$]{\resizebox{0.38\textwidth}{!}{
\begin{tikzpicture}

\definecolor{darkgray176}{RGB}{176,176,176}
\definecolor{darkorange25512714}{RGB}{255,127,14}
\definecolor{forestgreen4416044}{RGB}{44,160,44}
\definecolor{lightgray204}{RGB}{204,204,204}
\definecolor{steelblue31119180}{RGB}{31,119,180}

\begin{axis}[
legend cell align={left},
legend style={fill opacity=0.8, draw opacity=1, text opacity=1, draw=lightgray204},
tick align=outside,
tick pos=left,
x grid style={darkgray176},
xlabel={\large{$t$}},
ticklabel style = {font = \large},
xmin=-273.95, xmax=5774.95,
xtick style={color=black},
y grid style={darkgray176},
ylabel={\large{$|x(t)-A|$}},
ymin=-0.0338457741410796, ymax=0.710899378531655,
ytick style={color=black},
ytick={-0.2,0,0.2,0.4,0.6,0.8},
yticklabels={\ensuremath{-}0.1,0.0,0.1,0.2,0.3,0.4}
]
\addplot [very thick, , steelblue31119180]
table {%
1 0.664435328
5 0.666866986
10 0.654468561
30 0.51045059
50 0.334960781
100 0.0976701409999998
250 0.00232952300000001
500 3.8777000000073e-05
750 2.92980000000733e-05
1000 3.02160000000736e-05
1500 1.80000000000735e-05
2000 4.86360000000733e-05
2500 1.81530000000733e-05
3000 1.80320000000733e-05
3500 1.85550000000733e-05
4000 1.85590000000734e-05
5000 3.80800000000733e-05
5500 2.40800000000733e-05
};
\addlegendentry{$r = 0.5$}
\addplot [semithick, steelblue31119180, opacity=0.5, forget plot]
table {%
1 0.65414539912242
5 0.656686645862561
10 0.64424313056804
30 0.499643696114073
50 0.326067339816136
100 0.0938265104994871
250 0.00183189297598925
500 1.3826623364011e-05
750 1.36600765771233e-05
1000 1.82484060086667e-05
1500 1.80000000000735e-05
2000 6.2782531356042e-06
2500 1.79410588686413e-05
3000 1.79876724431727e-05
3500 1.79902801977114e-05
4000 1.77846529904614e-05
5000 1.0575090745417e-05
5500 1.86106008426084e-05
};
\addplot [semithick, steelblue31119180, opacity=0.5, forget plot]
table {%
1 0.67472525687758
5 0.677047326137439
10 0.664693991431961
30 0.521257483885927
50 0.343854222183865
100 0.101513771500513
250 0.00282715302401077
500 6.3727376636135e-05
750 4.49359234230232e-05
1000 4.21835939914805e-05
1500 1.80000000000735e-05
2000 9.09937468645424e-05
2500 1.83649411315054e-05
3000 1.80763275569739e-05
3500 1.91197198024351e-05
4000 1.93333470096854e-05
5000 6.55849092547297e-05
5500 2.95493991575382e-05
};
\addplot [very thick, dotted, , darkorange25512714]
table {%
1 0.657017058
5 0.649308873
10 0.643123325999999
30 0.620405551
50 0.601600497
100 0.574324333000001
250 0.471945589
500 0.336841675000002
750 0.239770186000002
1000 0.171970003000001
1500 0.0867282229999996
2000 0.0437892989999996
2500 0.027660992
3000 0.015489307
3500 0.006006042
4000 0.00540274
5000 0.00101051500000004
5500 0.00126339300000004
};
\addlegendentry{$r = 1.0$}
\addplot [semithick, darkorange25512714, opacity=0.5, forget plot]
table {%
1 0.646354946237972
5 0.638557232652819
10 0.632278048363123
30 0.609222825474667
50 0.590213018231343
100 0.562068467442088
250 0.457824536201659
500 0.322230877700488
750 0.225992245524094
1000 0.159598032463179
1500 0.0770969745342678
2000 0.0367877373425004
2500 0.0220114858618161
3000 0.011123009515406
3500 0.00364187835818511
4000 0.0028604534151522
5000 0.000193502633250575
5500 0.000314888455063228
};
\addplot [semithick, darkorange25512714, opacity=0.5, forget plot]
table {%
1 0.667679169762028
5 0.660060513347182
10 0.653968603636875
30 0.631588276525333
50 0.612987975768656
100 0.586580198557914
250 0.486066641798341
500 0.351452472299516
750 0.253548126475909
1000 0.184341973536824
1500 0.0963594714657315
2000 0.0507908606574989
2500 0.0333104981381839
3000 0.0198556044845939
3500 0.00837020564181489
4000 0.0079450265848478
5000 0.0018275273667495
5500 0.00221189754493686
};
\addplot [very thick, dashed, , forestgreen4416044]
table {%
1 0.664741522
5 0.656623288000001
10 0.627965958
30 0.406247378
50 0.212735444
100 0.028362472
250 2.79319999999995e-05
500 1.99979999999996e-05
750 1.99979999999996e-05
1000 1.99979999999996e-05
1500 1.99979999999996e-05
2000 1.99979999999996e-05
2500 1.99979999999996e-05
3000 1.99979999999996e-05
3500 1.99979999999996e-05
4000 1.99979999999996e-05
5000 1.99979999999996e-05
5500 1.99979999999996e-05
};
\addlegendentry{$r = 2.0$}
\addplot [semithick, forestgreen4416044, opacity=0.5, forget plot]
table {%
1 0.654113279981676
5 0.645748127835494
10 0.616848825422808
30 0.395709310762051
50 0.204165839160624
100 0.0253814911600156
250 2.56749634590856e-05
500 1.99952295276934e-05
750 1.99952295276934e-05
1000 1.99952295276934e-05
1500 1.99952295276934e-05
2000 1.99952295276934e-05
2500 1.99952295276934e-05
3000 1.99952295276934e-05
3500 1.99952295276934e-05
4000 1.99952295276934e-05
5000 1.99952295276934e-05
5500 1.99952295276934e-05
};
\addplot [semithick, forestgreen4416044, opacity=0.5, forget plot]
table {%
1 0.675369764018324
5 0.667498448164508
10 0.639083090577192
30 0.41678544523795
50 0.221305048839376
100 0.0313434528399844
250 3.01890365409134e-05
500 2.00007704723058e-05
750 2.00007704723058e-05
1000 2.00007704723058e-05
1500 2.00007704723058e-05
2000 2.00007704723058e-05
2500 2.00007704723058e-05
3000 2.00007704723058e-05
3500 2.00007704723058e-05
4000 2.00007704723058e-05
5000 2.00007704723058e-05
5500 2.00007704723058e-05
};
\end{axis}

\end{tikzpicture}\label{fig:pre_a5}}}\\
    \subfloat[$\alpha = 0.25$]{\resizebox{0.38\textwidth}{!}{
\begin{tikzpicture}

\definecolor{darkgray176}{RGB}{176,176,176}
\definecolor{darkorange25512714}{RGB}{255,127,14}
\definecolor{forestgreen4416044}{RGB}{44,160,44}
\definecolor{lightgray204}{RGB}{204,204,204}
\definecolor{steelblue31119180}{RGB}{31,119,180}

\begin{axis}[
legend cell align={left},
legend style={fill opacity=0.8, draw opacity=1, text opacity=1, draw=lightgray204},
tick align=outside,
tick pos=left,
x grid style={darkgray176},
xlabel={\large{$t$}},
ticklabel style = {font = \large},
xmin=-11.45, xmax=262.45,
xtick style={color=black},
y grid style={darkgray176},
ylabel={\large{$|x(t)-A|$}},
ymin=-0.0341801886750985, ymax=0.717821395507552,
ytick style={color=black},
ytick={-0.2,0,0.2,0.4,0.6,0.8},
yticklabels={\ensuremath{-}0.1,0.0,0.1,0.2,0.3,0.4}
]
\addplot [very thick,  , steelblue31119180]
table {%
1 0.672982641999999
5 0.514292267
10 0.311591177
30 0.0285357009999999
50 0.00393896100000007
100 0.000276454000000057
250 2.7820000000574e-06
};
\addlegendentry{$r = 0.5$}
\addplot [semithick, steelblue31119180, opacity=0.5, forget plot]
table {%
1 0.662325778682567
5 0.502619963143559
10 0.300161228211716
30 0.024482301310051
50 0.00204215772150707
100 7.86479008909512e-06
250 1.70151502195699e-06
};
\addplot [semithick, steelblue31119180, opacity=0.5, forget plot]
table {%
1 0.683639505317432
5 0.525964570856441
10 0.323021125788283
30 0.0325891006899488
50 0.00583576427849308
100 0.000545043209911019
250 3.8624849781578e-06
};
\addplot [very thick,  dotted,, darkorange25512714]
table {%
1 0.646888399000001
5 0.56067252
10 0.477887374999999
30 0.237265404
50 0.127087603
100 0.019778362
250 3.04200000002756e-06
};
\addlegendentry{$r = 1.0$}
\addplot [semithick, darkorange25512714, opacity=0.5, forget plot]
table {%
1 0.635801088760631
5 0.54802476593219
10 0.464147206638081
30 0.223908977987857
50 0.115853465717296
100 0.0151700486788111
250 2.99816789639689e-06
};
\addplot [semithick, darkorange25512714, opacity=0.5, forget plot]
table {%
1 0.657975709239371
5 0.57332027406781
10 0.491627543361917
30 0.250621830012144
50 0.138321740282704
100 0.024386675321189
250 3.08583210365823e-06
};
\addplot [very thick,  dashed,, forestgreen4416044]
table {%
1 0.642982613
5 0.427370308
10 0.201854046
30 0.00493608199999999
50 5.55639999999987e-05
100 3.95600000000133e-06
250 3.95600000000133e-06
};
\addlegendentry{$r = 2.0$}
\addplot [semithick, forestgreen4416044, opacity=0.5, forget plot]
table {%
1 0.631944061344016
5 0.415569917659886
10 0.191600628010981
30 0.00323019222984486
50 1.29340661015233e-05
100 3.94314263902831e-06
250 3.94314263902831e-06
};
\addplot [semithick, forestgreen4416044, opacity=0.5, forget plot]
table {%
1 0.654021164655984
5 0.439170698340113
10 0.212107463989019
30 0.00664197177015513
50 9.81939338984741e-05
100 3.96885736097435e-06
250 3.96885736097435e-06
};
\end{axis}

\end{tikzpicture}\label{fig:pre_a25}}}\\
    \subfloat[$\alpha = 0.50$]{\resizebox{0.38\textwidth}{!}{
\begin{tikzpicture}

\definecolor{darkgray176}{RGB}{176,176,176}
\definecolor{darkorange25512714}{RGB}{255,127,14}
\definecolor{forestgreen4416044}{RGB}{44,160,44}
\definecolor{lightgray204}{RGB}{204,204,204}
\definecolor{steelblue31119180}{RGB}{31,119,180}

\begin{axis}[
legend cell align={left},
legend style={fill opacity=0.8, draw opacity=1, text opacity=1, draw=lightgray204},
tick align=outside,
tick pos=left,
x grid style={darkgray176},
xlabel={\large{$t$}},
ticklabel style = {font = \large},
xmin=-3.95, xmax=104.95,
xtick style={color=black},
y grid style={darkgray176},
ylabel={\large{$|x(t)-A|$}},
ymin=-0.0307282956407477, ymax=0.645296593058924,
ytick style={color=black}
]
\addplot [very thick, , steelblue31119180]
table {%
1 0.588812678
5 0.273243922
10 0.0901545299999998
30 0.00154626200000005
50 2.05600000005624e-06
100 1.96400000005648e-06
};
\addlegendentry{$r = 0.5$}
\addplot [semithick, steelblue31119180, opacity=0.5, forget plot]
table {%
1 0.576227737187171
5 0.259945608774189
10 0.0810362994519519
30 0.00028843580502065
50 1.95441162881648e-06
100 1.9523463206715e-06
};
\addplot [semithick, steelblue31119180, opacity=0.5, forget plot]
table {%
1 0.60139761881283
5 0.286542235225811
10 0.0992727605480477
30 0.00280408819497945
50 2.15758837129599e-06
100 1.97565367944145e-06
};
\addplot [very thick, dotted,, darkorange25512714]
table {%
1 0.601525304
5 0.357463373
10 0.192258131
30 0.016451401
50 0.00102462300000003
100 9.62000000027663e-07
};
\addlegendentry{$r = 1.0$}
\addplot [semithick, darkorange25512714, opacity=0.5, forget plot]
table {%
1 0.588482418972879
5 0.342434514165311
10 0.178684112837545
30 0.0120034737217388
50 1.43948592955204e-05
100 9.18204722143966e-07
};
\addplot [semithick, darkorange25512714, opacity=0.5, forget plot]
table {%
1 0.614568189027121
5 0.372492231834689
10 0.205832149162455
30 0.0208993282782613
50 0.00203485114070453
100 1.00579527791136e-06
};
\addplot [very thick, dashed,, forestgreen4416044]
table {%
1 0.557420877
5 0.190928592
10 0.0399130979999999
30 1.17400000000371e-06
50 1.30000000003738e-07
100 1.30000000003738e-07
};
\addlegendentry{$r = 2.0$}
\addplot [semithick, forestgreen4416044, opacity=0.5, forget plot]
table {%
1 0.544398543303147
5 0.178417148461613
10 0.0334410392759222
30 1.10315996053675e-07
50 1.08391055558744e-07
100 1.08391055558744e-07
};
\addplot [semithick, forestgreen4416044, opacity=0.5, forget plot]
table {%
1 0.570443210696853
5 0.203440035538386
10 0.0463851567240776
30 2.23768400395375e-06
50 1.51608944448733e-07
100 1.51608944448733e-07
};
\end{axis}

\end{tikzpicture}\label{fig:pre_a50}}}
    \caption{Average minimum Manhattan distance to $A$ over time for various $\alpha$ and $r.$}
\end{figure}
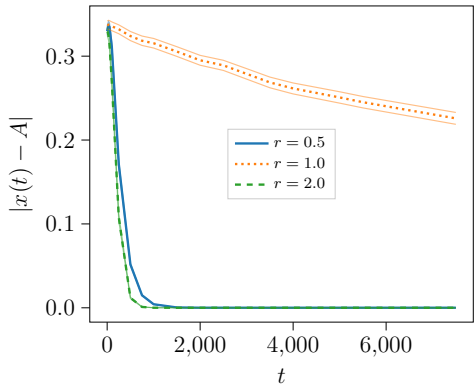
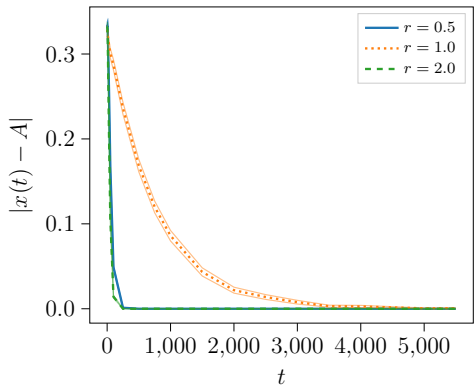
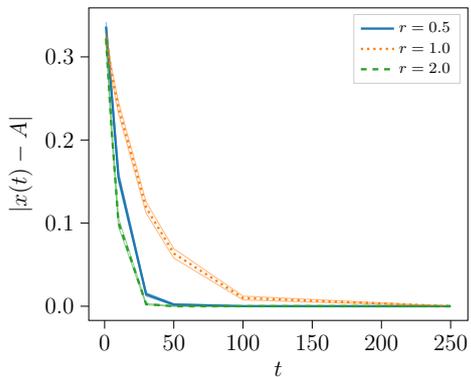
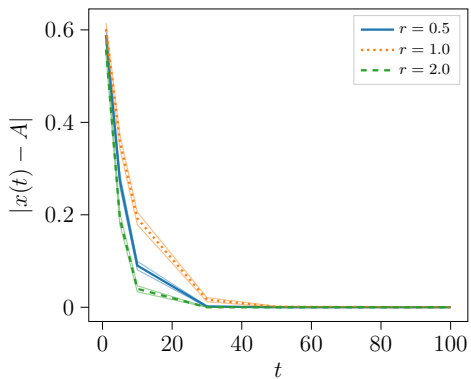

We parameterize $F$ by $r\in [0.5,2]$ and take only functions of the form $F(x)=x^r$. In particular note that $F(x) = \sqrt{x}$ and $F(x)=x^2$ are included in this parameterization. The two agents are initialized with beliefs chosen uniformly at random over the belief space $(0,1)$. Thus $\bm{x}(0)\sim \mathcal{U}_{[0,1]}^2$. In order to determine how many time steps ought to be simulated to allow process to be absorbed in one of the two steady states we run a preliminary simulation for the values $\alpha = \{0.01, 0.05, 0.25, 0.5\}$ and $r = \{0.5,1,2\}$. We run 1000 iterations for 10\,000 times steps keeping track of the average Manhattan distance to the nearest steady state over time. The resulting 95\% confidence bounds of the average Manhattan distance to $A$ is depicted in Figures~\ref{fig:pre_a1}--\ref{fig:pre_a50}.

In Figure~\ref{fig:pre_a1} we see that assuming convergence at some time $t<10\,000$ when $r=1$ and $\alpha\leq 0.01$ is devious and should be avoided. For $\alpha = 0.05, 0.25, 0.5$ we see that 5\,000, 250 and 100 rounds respectively are enough to be 95\% confident that the process has indeed been absorbed in a steady state. For the simulation study, we run the simulation for 1000 time steps when $\alpha \geq 0.25$, and for 10\,000 time steps for $\alpha < 0.25$. In particular we run simulations for $\alpha\in \{0.05,0.1,\allowbreak 0.2,\allowbreak 0.3,\allowbreak 0.4,0.5\}$ for values of $r\in \{0.5, 0.6,\allowbreak 0.7,\allowbreak 0.8,\allowbreak 0.9,\allowbreak 0.95,\allowbreak 1, 1.05,\allowbreak 1.1,\allowbreak 1.25,\allowbreak 1.40, 1.55, 2\}$.

\begin{figure}
    \centering
    \subfloat[$p_T(\bm{x})$, where $\bm{x}\sim \mathcal{U}_{0,1}^2$]{\includegraphics[width = 0.49\textwidth]{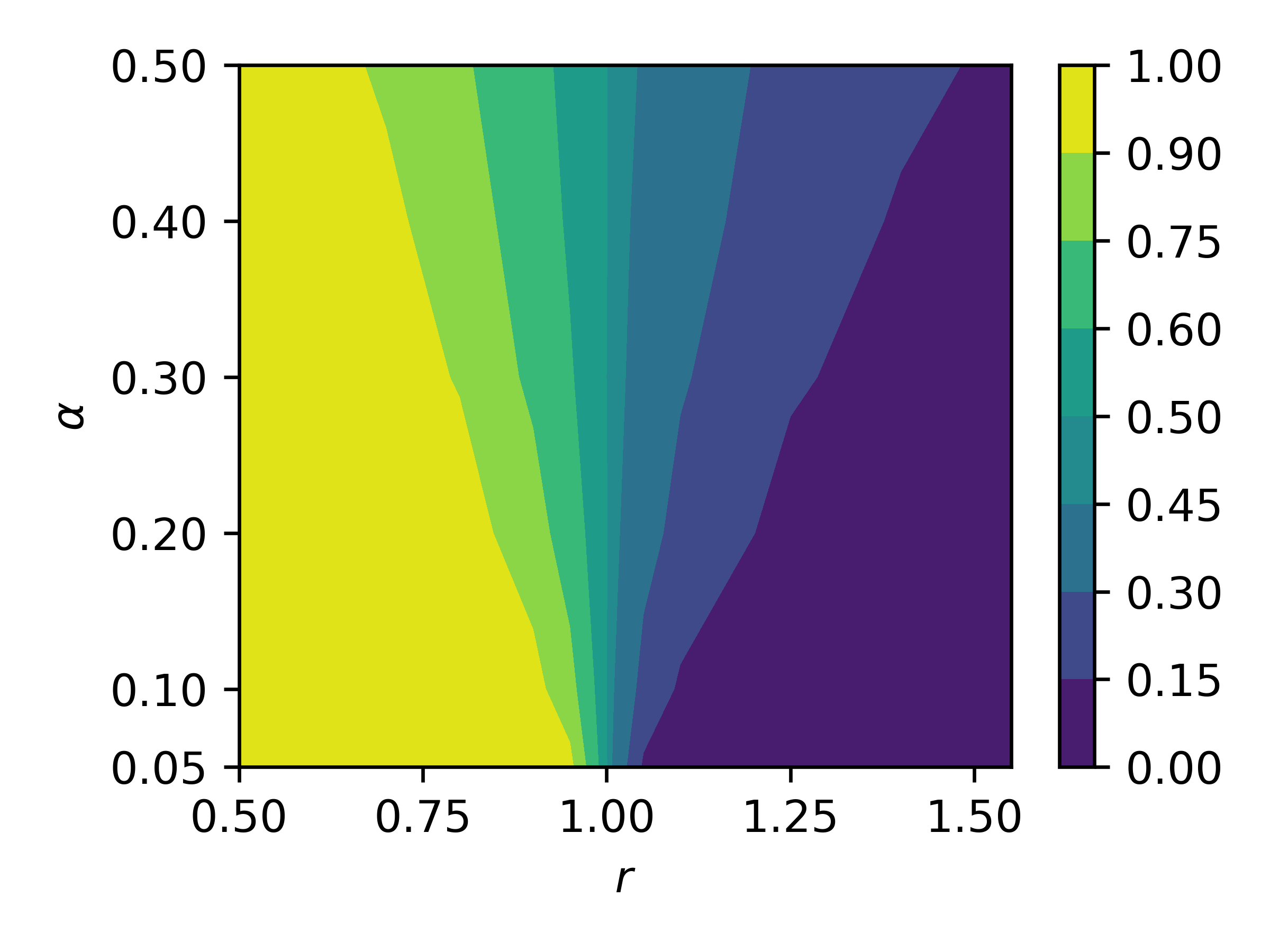}\label{fig:HM_abs_1}}
    \\
    \subfloat[Time until absorption]{\includegraphics[width = 0.49\textwidth]{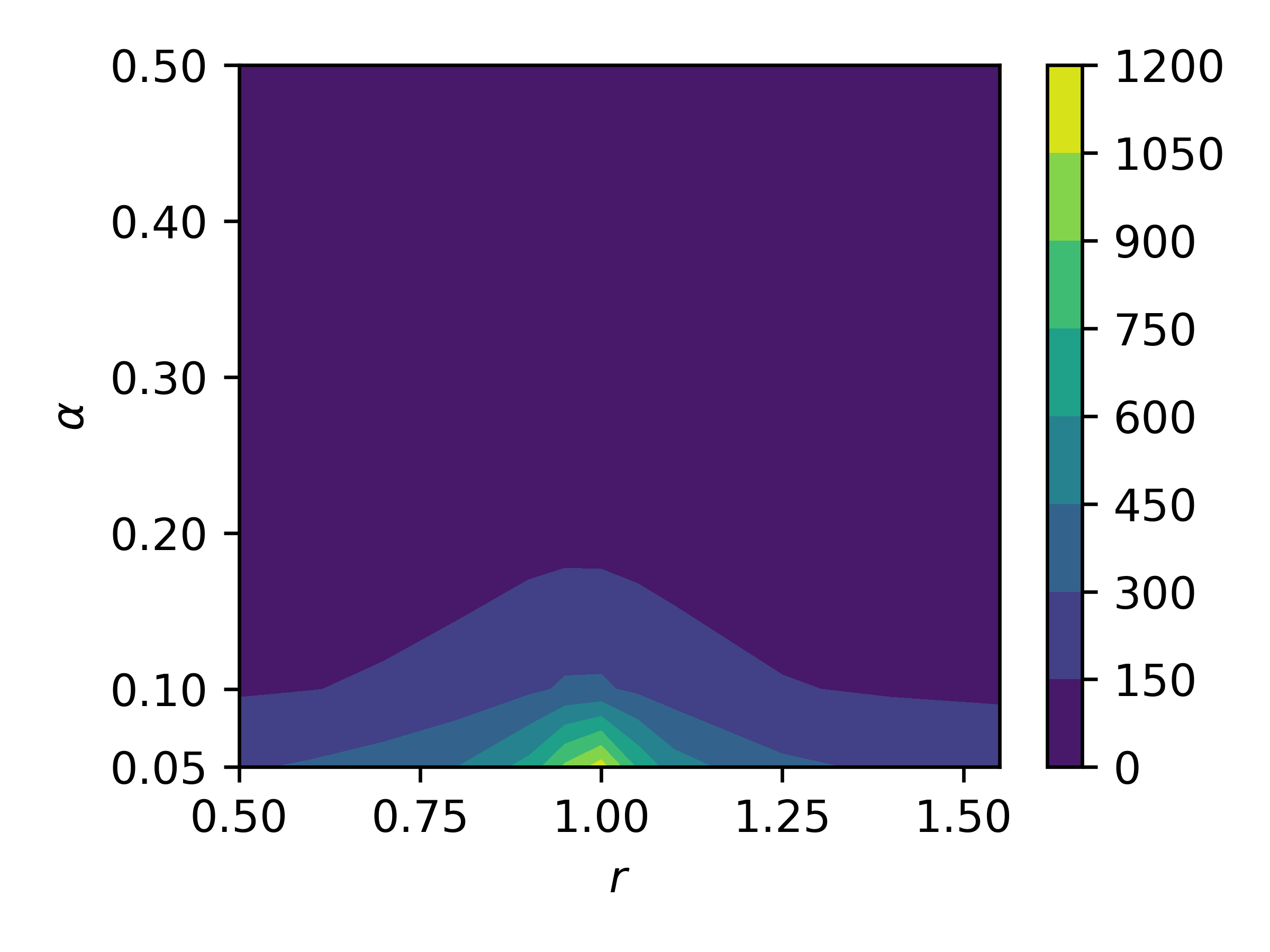}\label{fig:HM_abs_time}}
    \caption{Heatmaps of $p_T(\bm{x}(0))$ and the time it takes to be absorbed at either corner. We see the transition between being absorbed in $(1,1)$ with probability 1 when $r<1$ and probability zero when $r>1$. A greater $\alpha$ has the effect of smoothing this transition. We also see that increasing $\alpha$ has a hastening effect on the time to absorption.}
\end{figure}

We show the probability of being absorbed in the always-trust corner of the state space in Figure~\ref{fig:HM_abs_1}. We notice that increasing $\alpha$ has the effect of smoothing the transition between being absorbed in $(0,0)$ and in $(1,1)$. Conversely, decreasing $\alpha$ results in a sharper transition between being absorbed in $(0,0)$ with probability one when $r>1$ and absorbed in $(1,1)$ with probability one when $r<1$. The following conjecture elaborates on this.

\begin{conjecture}[Phase transition at $r=1$ as $\alpha\to 0$]\label{con:phase-transition}
    As $\alpha\to 0$ the $p_T(\bm{x}) = 1$ when $r<1$ and $p_T(\bm{x}) = 0$ when $r>1$ for all $\bm{x}\in [0,1]^N \setminus \{\bm{0}^N,\bm{1}^N\}$. 
\end{conjecture}

The machinery used in the proofs requires $\epsilon<\alpha$. This provides clear criteria for events that lead the population to `jump out' of the $\epsilon$-balls around zero and one. It also ensures that the population can converge to an $\epsilon$-ball around zero or one in finite time (see division by $\log (\bar{\alpha}) = \log(1-\alpha)$ in (\ref{eq:k_finite}) for example). New techniques would thus have to be applied in order to prove Conjecture~\ref{con:phase-transition}. Intuitively however, as the learning rate $\alpha$ goes to zero, the size of the steps agents make when they adjust their belief shrink. This seems to imply that an infinite number of steps might be needed but it is only their direction that matters and so movement towards the expectation (to zero when $r>1$ and toward one when $r<1$) will eventually come to pass. 

Phase transitions are quite common in evolutionary game theory, with some recent results in~\cite{Zheng2017, Duong2020, Zeng2022}. Duong and Han~\cite{Duong2020} also conjecture a phase transition (regarding the expected number of equilibria) whose proof is beyond the standard techniques for phase transitions in the replicator dynamics. Zheng \textit{et al.}\,\cite{Zheng2017} and Zeng \textit{et al.}\,\cite{Zeng2022} present phase transitions for imitation learning dynamics, which they observe in simulations. Though beyond the scope of this paper, Conjecture~\ref{con:phase-transition} could be investigated in future work by means of an extensive simulation study.

In Figure~\ref{fig:HM_abs_time} we show the time average time until the simulation entered $A$ for the first time. By choosing $\epsilon = 10^{-5}$, we suggest that the time it takes to enter $A = [0,\epsilon]^2 \cup [1-\epsilon,1]^2$ is proportional to the time it takes to also be absorbed in $A.$ Supposing this is the case, then we see that a greater $\alpha$ has the effect of speeding up the dynamics. This is sensible as the step sizes made by the process are bigger for a bigger $\alpha.$ We also see that the closer $r$ is to $1$, the longer the dynamics take. This is explainable by the fact that with a greater $r$, while the process is still far from $A$, the probability of steps toward the center are still fairly likely. In short, smaller $|1-r|$ implies that the process spends more time in the center of the state space, while lower $\alpha$ means that the process is moving to the corners at a slower pace.

\begin{figure*}
    \centering
    \subfloat[$p_T$ for $N=5$]{\includegraphics[width = 0.49\textwidth]{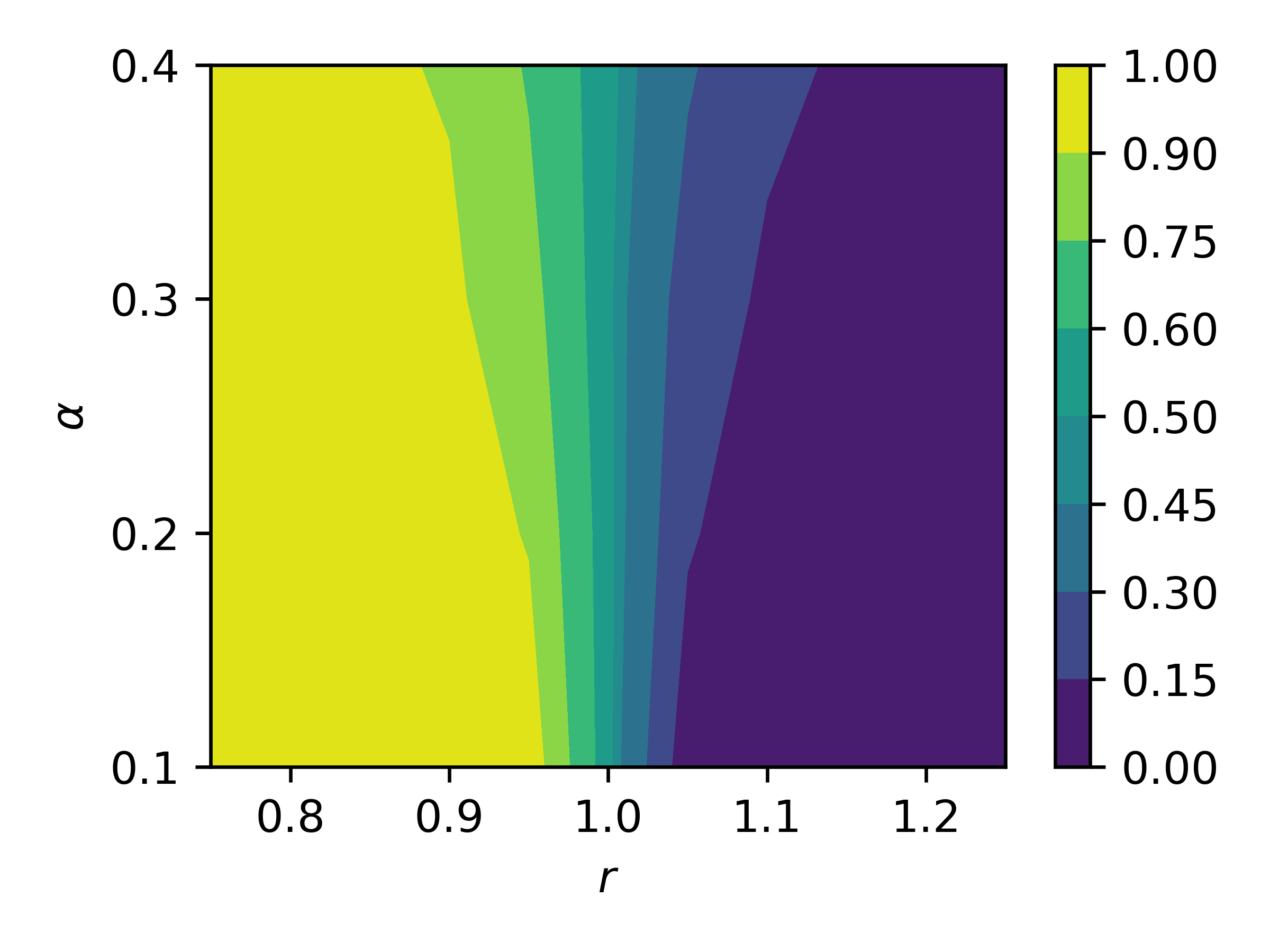}\label{fig:HM_abs_5}}
    \subfloat[$p_T$ for $N=10$]{\includegraphics[width = 0.49\textwidth]{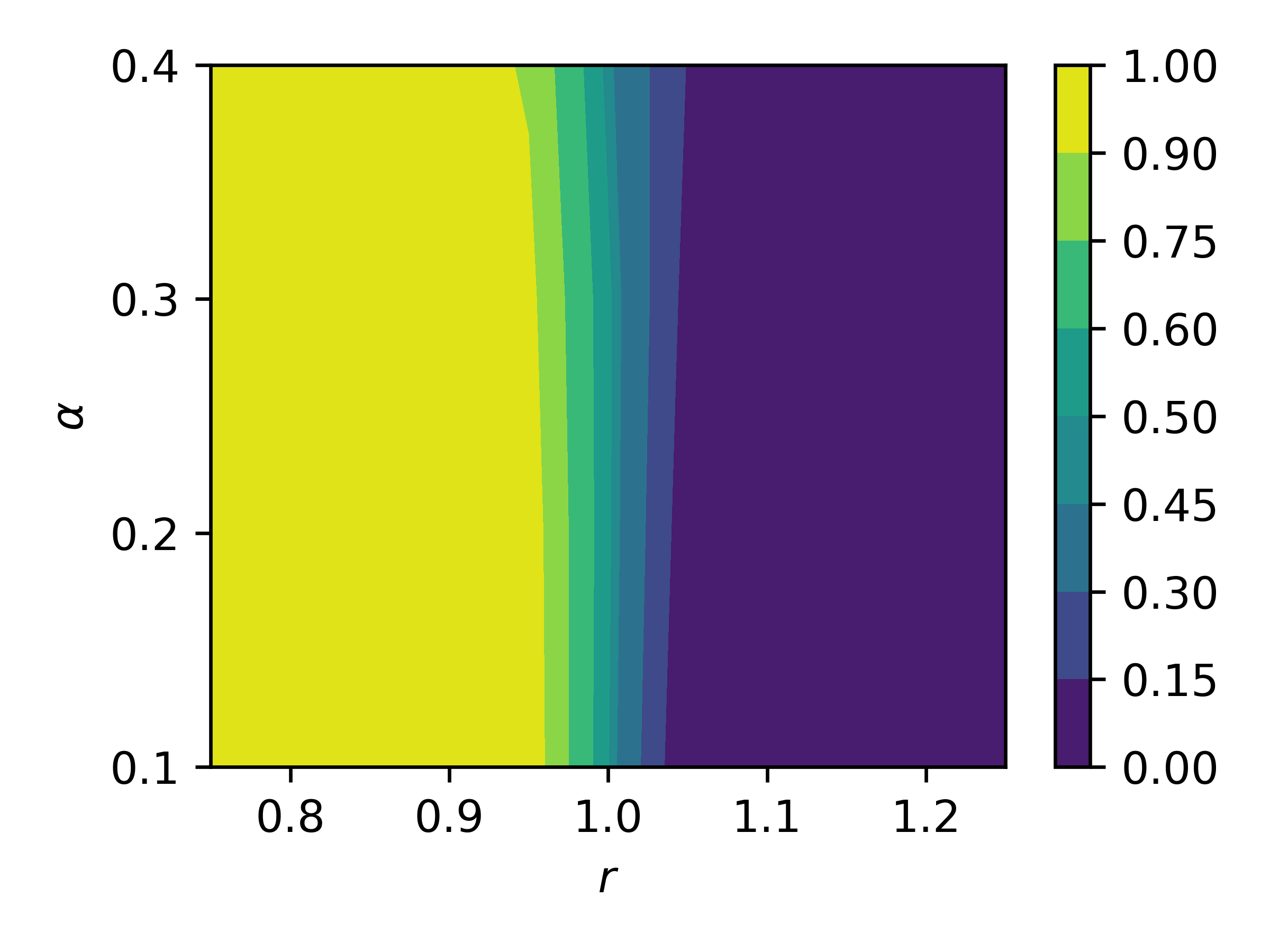}\label{fig:HM_abs_10}}\\
    \subfloat[Time $N=5$]{\includegraphics[width = 0.49\textwidth]{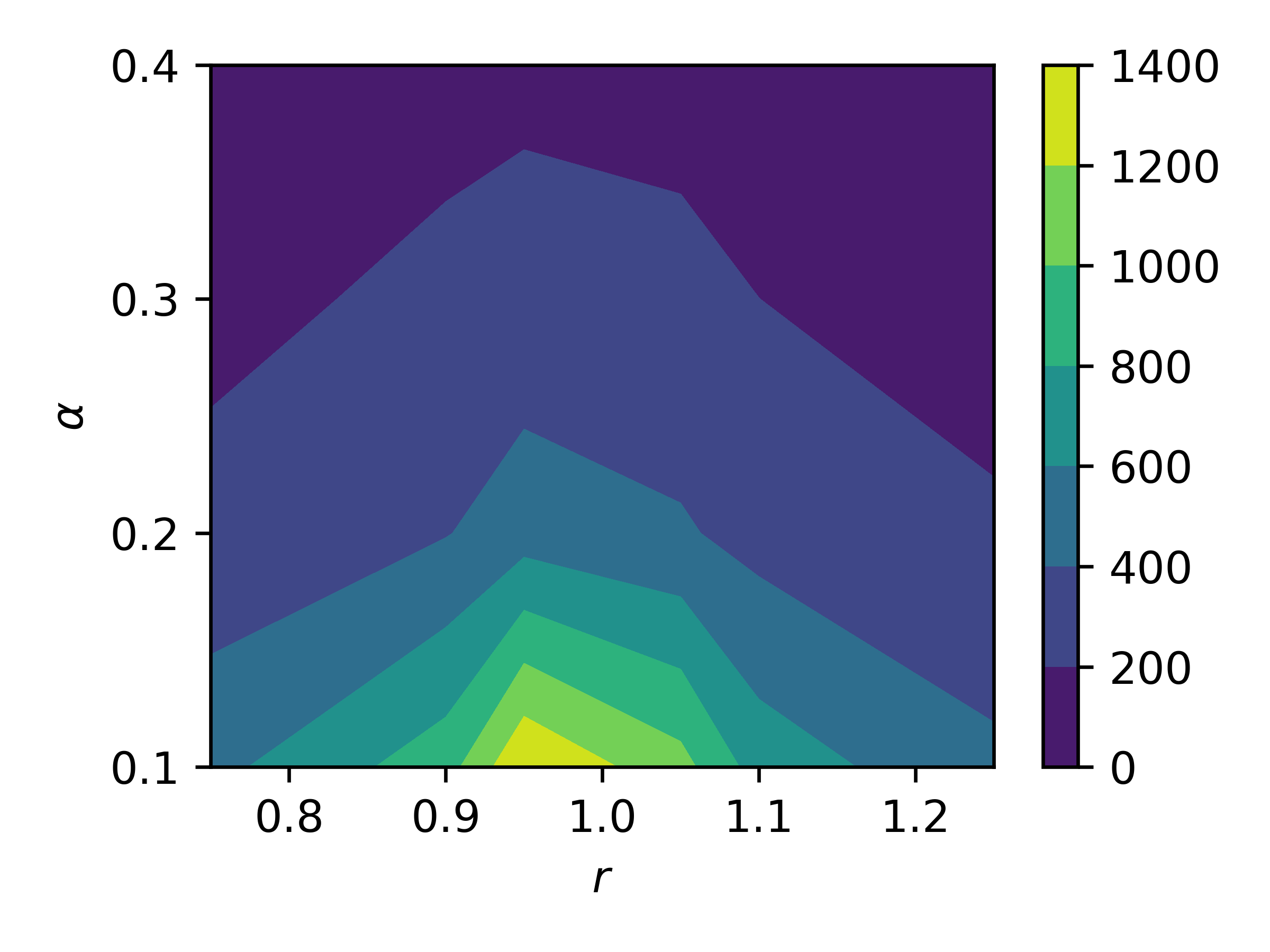}\label{fig:HM_time_5}}
    \subfloat[Time $N=10$]{\includegraphics[width = 0.49\textwidth]{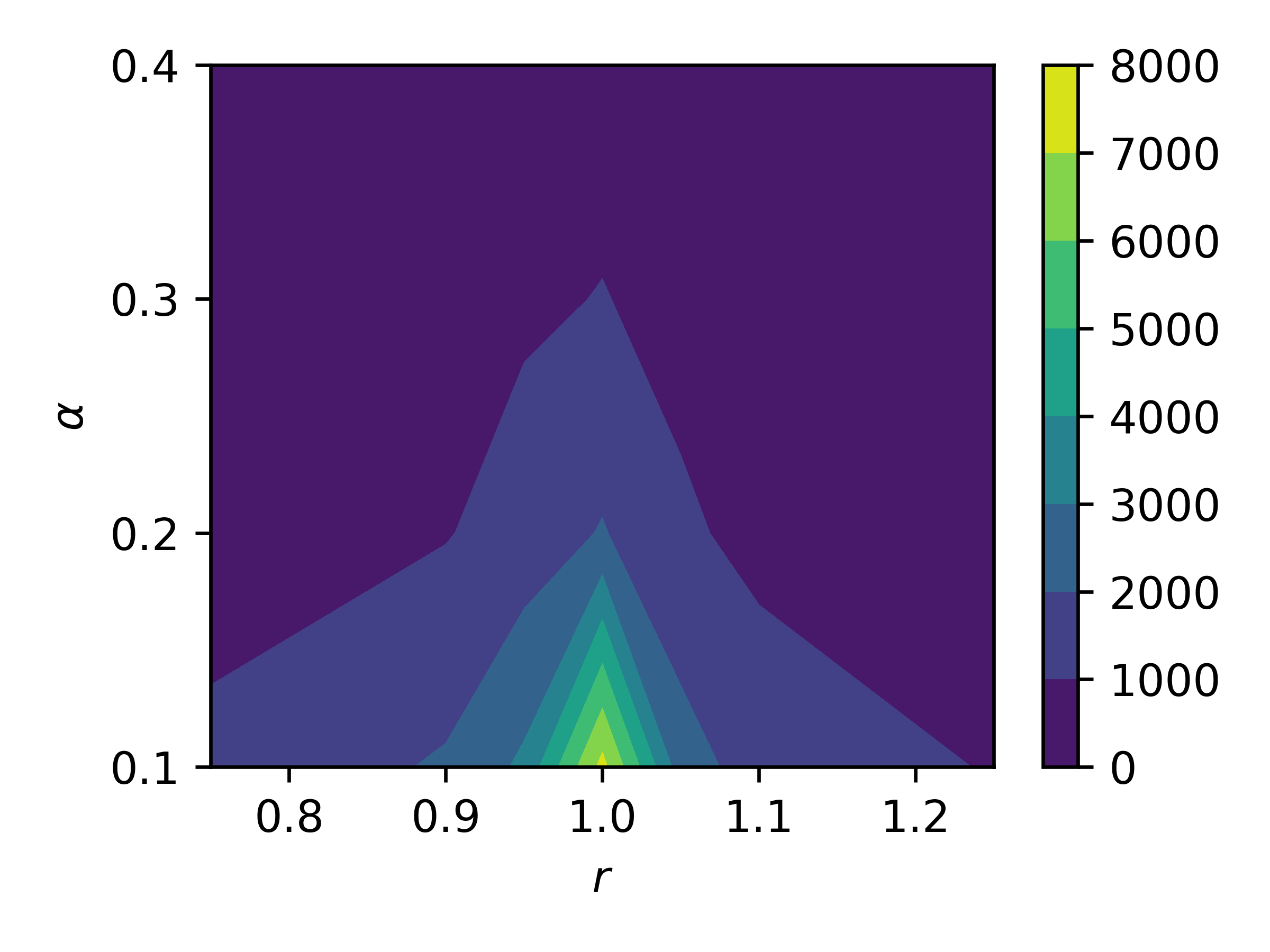}\label{fig:HM_time_10}}
    \caption{Heatmaps of $p_T(\bm{x}(0))$ and the time it takes to be absorbed at either corner. Increasing the number of agents in the population dampens the effect of the model learning rate $\alpha$ on the outcome or sharpening the effect of the payoff distribution parameter $r$.}
\end{figure*}

In order to show how the behavior changes as the population size is increased, we present simulations for $N=5$ and $N=10$. The resulting probability of absorption at $\bm{1}^N$ is depicted in heatmap-format in Figure~\ref{fig:HM_abs_5}--\ref{fig:HM_abs_10}. For $N=5$, in Figure~\ref{fig:HM_abs_5} we already see that the effect of $\alpha$ observed for $N=2$ is less pronounced, though still present. Furthermore, for $N=10$ in Figure~\ref{fig:HM_abs_10} we see that $\alpha$ almost plays no role in terms of the relative absorption probabilities. Thus as $N$ increases we suspect that the steady state in which the process will be absorbed depends more on the cdf $F$ and less on the learning dynamics of the individual agents. 

The time to absorption for the simulation runs with $N=5$ and $N=10$ are shown in Figure~\ref{fig:HM_time_5} and~\ref{fig:HM_time_10} respectively. We see that, qualitatively, the interplay between $\alpha$ and $r$ on the time to absorption is the same. That is, bigger $\alpha$ and $|1-r|$, speed up the dynamics.

\section{Conclusion}
\label{Conclusion}
We modeled the problem of trusting strangers in society as game in which agents of a population are randomly matched, two per round, and tasked with a coordination game with random payoffs. These agents are endowed with a learning procedure by which they update their belief on the probability that a random stranger would trust using the exponential moving average of their passed observations. We have shown for that for any finite population of size $N$, with mild conditions on the cdf of the payoff parameters, and a constant learning rate (or memory) $\alpha \in (0,1)$ the process is absorbed with probability $1$ at one of the two steady states: always-trust, and always-doubt. This result is not immediately obvious because looking at the expected change of the process there are $F$ for which it would seem that all agents believing $\bm{x}=[0.5]^N$ is an attracting steady state. 

Furthermore, we analyzed the effect of $F$, $\alpha$ and $N$ by simulating the model with parameterized $F(x) = x^r$ for $r\in [0.5,2]$ and $\alpha \in [0.05,0.5]$. The results of this simulation show that
\begin{itemize}
    \item decreasing $\alpha$ exaggerates the effect $r$
    \item increasing $N$ decreases the effect of  $\alpha$
\end{itemize} 

We conjecture that as $\alpha\to 0$ there will emerge a sharp phase transition at $r=1$, such that the process is absorbed in the always-trust steady state with probability 1 when $r<1$ and with probability 0 when $r>1.$ 

A similar phase transition might occur as the population size is increased. 
A broad take away of this model is that differences in trust between populations, might simply boil down to chance; the process by which a population learns on which action to coordinate is random and is not necessarily fully determined by the nature of the game ($F$), the size of the population ($N$) or the rate of learning ($\alpha$). However, in the case of large populations our simulations suggest that the context of the interaction (distributions of the payoff parameters) will play a role in determining whether or not they end up with long-term trust or long-term doubting. In terms of the numerical simulation, the probability of converging to the always-trust steady state decreases (increases) as the population grows, for $r>1$ ($r<1$). This implies that the effect of increasing population is not determined by the type of game (like for the one-shot $N$-player public goods and prisoner's dilemma games shown in~\cite{Barcelo2015}) but really the specific payoff structure. For instance, our model corresponds to an all-or-nothing version of the 2 player public goods game when $F$ is a step function with one step from zero to one. Players either pay one (when $A_k=T)$ or zero (when $A_k = D$) while the reward $\gamma \in \mathbb{R}$ is only paid out if \textit{both} players payed the cost. Then whether or not a larger population benefits one or the other steady state depends not on the fact that it is a public goods game but specifically where the step function steps from zero to one. Specifically the location of this step corresponds to $1/\gamma$.

As possible answer to our research question, we see that societies may be low or high trust largely due to the context of the interactions within them. The context being the distributions of the payoffs in the `coordination game' the people are playing with one another. Adjusting the payoffs of the social interactions might thus be the most effective way to foster more trust in a population. Chance plays a bigger role in smaller groups than with large ones. Future research could be to investigate whether a large population could be modeled by a lose grouping of smaller cliques and that the average trusting behavior of the population is an aggregation of these cliques. In this case chance may indeed play a big role in determining the amount of trust on a clique level, which indirectly effects the trust in the population as a whole. 

The model we consider assumes a population structure in which all agents are connected with one another. Although the result of Theorem~\ref{thm:n_abs} should also hold for any other network topology as long as this is connected, it might interesting to look for the possibility of pseudo steady states in networks with a strong community structures. It could also be interesting to characterize a mapping between the structure of the graph defining the interactions and the rate of convergence to a steady state. Other future work could focus on extending the convergence result (or proving its opposite) in the case of Bayesian learning or another simple learning rule that the agents might employ.

Our model is such that if there exists even one agent who always trusts regardless of the actions taken by the rest of the population, the population will converge to the trust action with probability one (because convergence to always doubt is impossible). This is in contrast to Wang and Sun~\cite{Wang2023} in whose model the presence of `zealous' always cooperators does not always promote cooperation. However, in our model the converse is also true: One constant doubter forces the always doubt equilibrium. Studying the transient dynamics of our model with both constant trusters and constant doubters might lead to results with oscillations between trusting and doubting as in~\cite{Szolnoki2016}.

Another line of future work could do away with the assumption of selfish rationality and instead have agents using some form of moral preferences showcased in \cite{Joireman1996, Capraro2021}. For example agents who are interested in the welfare of both players in the game. A facet of trust we have not explored here has to do with believing what has been said by others. Recently, the evolution of honesty in the sender-receiver game has been studied under imitation of better-performing strategies by means of the Monte Carlo method~\cite{Capraro2019,Capraro2020,Kumar2021}. In particular, agents in these studies copy the behavior of other agents at probability proportional to the difference between their payoffs. It may be fruitful to adapt the approach we have taken here to study the evolution of honesty in the sender-receiver game under experience based learning instead of imitation. In this case it may be necessary for agents to have two beliefs: $x$ (likelihood of being believed) and $y$ (likelihood of others being honest). In particular it may be interested to investigate under which conditions the golden rule (lie and disbelieve or be honest and believe) emerges.

\section*{Acknowledgments}
The authors report there are no competing interests to declare.

\noindent  This research was supported by the European Union’s Horizon 2020 research and innovation programme under the Marie Skłodowska-Curie grant agreement no. 945045, and by the NWO Gravitation project NETWORKS under grant no. 024.002.003. \includegraphics[height=1em]{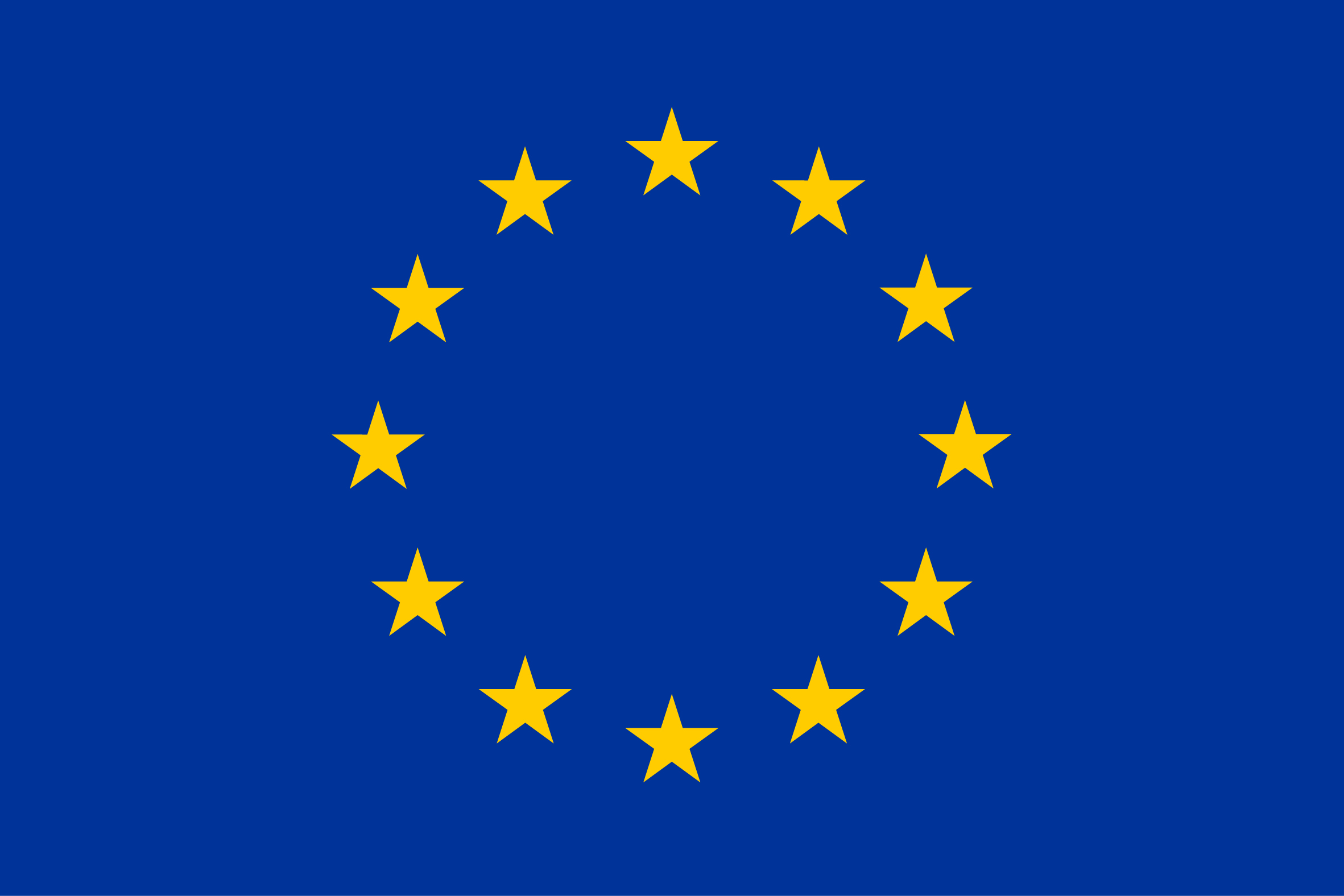}

\bibliographystyle{ieeetr}
\bibliography{Biblio.bib}

\end{document}